\documentclass[a4paper,11pt]{article}
\pdfoutput=1 

\usepackage{jinstpub} 

\title{\boldmath Magnetic field simulations and measurements on the 
mini-ICAL detector }

\author[a,b,1]{Honey Khindri,\note{Corresponding author.}}
\author[c]{B. Satyanarayana,}
\author[b]{D. Indumathi,}
\author[b]{V.M. Datar,}
\author[c]{R. Shinde,}
\author[d]{N. Dalal,}
\author[d]{S. Prabhakar,}
\author[d]{S. Ajith}

\affiliation[a]{Homi Bhabha National Institute,\\
Anushakti Nagar, Mumbai 400094, India}
\affiliation[b]{The Institute of Mathematical Sciences,\\
Taramani, Chennai 600113, India}
\affiliation[c]{Tata Institute of Fundamental Research,\\
Homi Bhabha Road, Mumbai 400005, India}
\affiliation[d]{Bhabha Atomic Research Centre,\\
Mumbai 400085, India}

\emailAdd{honey1661988@gmail.com}

\abstract{The ICAL (Iron Calorimeter) is a 51 kTon magnetized detector
proposed by the INO collaboration. It is designed to detect muons with
energies in the 1--20 GeV range.  A magnetic field of $\sim$ 1.5 T in the
ICAL detector will be generated by passing a DC current through suitable copper
coils. This will enable it to distinguish between $\mu^-$ and $\mu^+$
that will be generated from the interaction of atmospheric $\nu_{\mu}$
and $\overline{\nu}_\mu$ with iron. This will help in resolving the open
question of mass ordering in the neutrino sector. Apart from charge
identification, the magnetic field will be used to reconstruct the
muon momentum (direction and magnitude). Therefore it is important to
know the magnetic field in the detector as accurately as possible. We
present here an (indirect) measurement of the magnetic field in the 85
ton prototype mini-ICAL detector working in Madurai, Tamil Nadu, for
different coil currents.  A detailed 3-D finite element simulation was
done for the mini-ICAL geometry using Infolytica MagNet software and the
magnetic field was computed for different coil currents. This paper presents,
for the first time, a comparison of the magnetic field measured in the
air gaps with the simulated magnetic field, to validate the simulation
using real time data. Using the simulations the magnetic
field inside the iron is estimated.}

\keywords{Normal-conducting magnet, Neutrino detectors}

\arxivnumber{} 

\begin{document}
\maketitle
\flushbottom

\section{Introduction and Motivation}
\label{sec:intro}

The Iron Calorimeter (ICAL) detector is a 51 kTon magnetized detector
proposed by the India-based Neutrino Observatory (INO) collaboration
\cite{ino:report}. It is designed to study atmospheric $\nu_{\mu}$
and $\overline{\nu}_\mu$ and explore the answer to some of the open
questions in the neutrino sector such as the neutrino mass ordering,
and make a precision measurement of some of the neutrino oscillation
parameters in the 2--3 sector. ICAL is optimized to detect both $\mu^-$
and $\mu^+$, in the energy range of 1-20 GeV generated by the charged current
(CC) interactions of $\nu_{\mu}$ and $\overline{\nu}_\mu$ respectively
with the iron. It is magnetized with sets of copper coils to achieve
a maximum magnetic field, B$_{max}$ $\sim$ 1.5 T. The presence of the
magnetic field will enable the ICAL detector to distinguish between
$\mu^-$ and $\mu^+$, and hence $\nu_{\mu}$ and $\overline{\nu}_\mu$
events; this can lead to resolving the mass ordering issue in the
neutrino sector making use of Earth matter effects on the neutrinos
and anti-neutrinos \cite{indu:murthy}. Apart from the separation of
charged particles, the magnetic field will be used to reconstruct the
muon momentum, both its magnitude and direction. Therefore the
magnetic field is one of the key aspects of the ICAL detector.

ICAL will consist of three modules, each having 151 layers of 56 mm thick
iron plates and 150 layers of RPCs placed in the 40 mm air gap between
two iron layers. Each module will be 16 m$\times 16$ m $\times 14.5$
m in length, width and height respectively; hence the total dimensions of
the ICAL will be 48 m$\times$ 16 m $\times$ 14.5 m. Each module will
be magnetized using 4 sets of copper coils with maximum magnetic field
of 1.5 T and more than 90$\%$ of the ICAL volume will experience more
than 1 T magnetic field.

Since ICAL detector is yet to be built, simulation studies have
been performed to explore the physics potential of the ICAL detector
\cite{ino:report}. These studies were done using the GEANT4 simulation
package \cite{geant}. A magnetic field map was generated for the ICAL
\cite{sp:bhr} using Infolytica MagNet software \cite{magnet6} for use
as input in the GEANT4 simulation. This map needs to be validated by
suitable measurements. This paper presents the first study of a
comparison between the measured and simulated magnetic field, not in the
main ICAL detector, but its scaled prototype, the mini-ICAL.

The mini-ICAL is an 85 ton magnetized (B$_{max} \sim 1.5$ T) prototype
detector built to study challenges while building and running the main
ICAL detector.  One of the aims of mini-ICAL detector is to validate the
magnetic field map generated using simulations with the real-time data
of the detector. Since there is no direct method to measure the magnetic
field inside iron, an indirect approach is used. Small gaps of specific
widths are introduced between sheets of iron that make up a single layer
of mini-ICAL.  The magnetic field is measured in these gaps between the
iron plates and compared with the simulated magnetic field. Results
of such a measurement for a single current of 500 A in the coil were
reported in Ref.~\cite{Khindri:2022elz}, \cite{hk:honey}. In this paper, a detailed study
of the measured magnetic field in mini-ICAL for different coil currents
from 500--900 A is reported. In addition, a detailed simulation of the
magnetic field for the mini-ICAL is performed using Infolytica MagNet
7 \cite{magnet6} software which uses 3-D finite element methods to
calculate the magnetic field.  A comparison between the simulated and
measured magnetic fields is presented in this paper and an estimation
of the magnetic field inside iron is made.

Section 2 contains details of the mini-ICAL geometry and a description of
the gaps at which the magnetic field measurements have been made. The
procedure involved and instruments used to measure the magnetic field in
the gaps has already been presented in Ref.~\cite{Khindri:2022elz} where
results for a coil current of 500 A were presented. The same method is
used in the current study and is described again here for completeness.
Section 3 contains the details on the measurement of the magnetic field
in mini-ICAL at different gaps for different currents in the coil. A
detailed simulation study listing the various inputs and parameters
used, and how the change in gap width affects the magnetic field
estimation is presented in Section 4. Finally, a comparison is made
between the measured and simulated fields in Section 5. While complete
agreement is not obtained between the two, Section 6 discusses possible
reasons for the deviations while concluding the paper.

\section{The mini-ICAL detector geometry}

\begin{figure}[btp]
  \centering
  \begin{minipage}[b]{0.49\textwidth}
    \includegraphics[width=\textwidth, height=0.79\textwidth ]{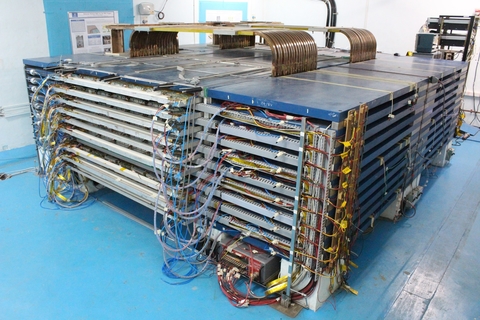}
    \caption{The {\sf mini-ICAL} detector with 11 iron layers. The
    upper portion of copper coils and slots are clearly visible.}
    \label{miniICAL}
  \end{minipage}
  \hfill
  \begin{minipage}[b]{0.49\textwidth}
    \includegraphics[width=\textwidth]{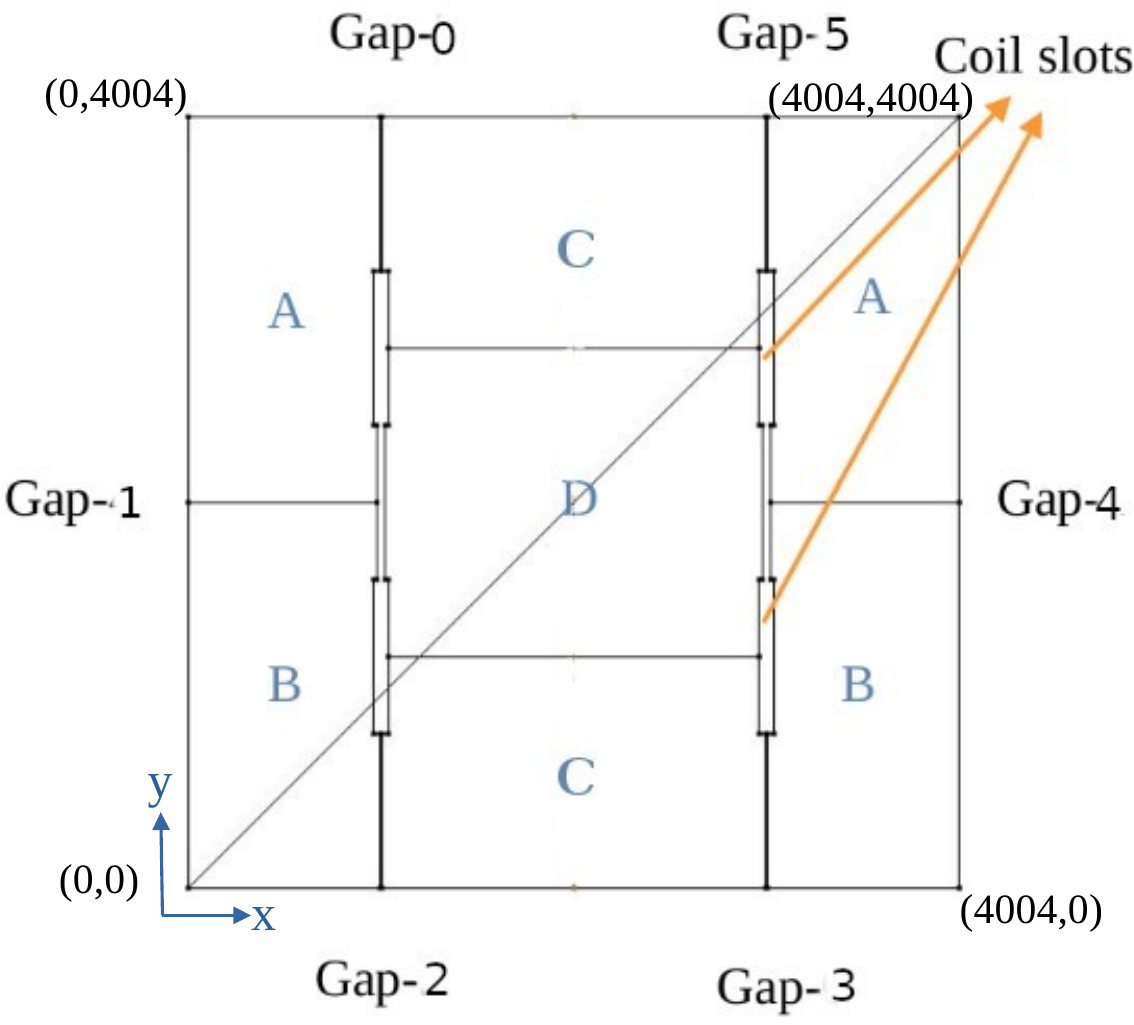}
    \caption{Schematic of top view of one {\sf mini-ICAL} layer showing the
    plates A, B, C, D and the gaps 0--5.}
    \label{final_mini_ical_up}
  \end{minipage}
\end{figure}

The 85 ton prototype detector mini-ICAL (Fig.~\ref{miniICAL}) consists
of 11 layers (numbered from 1--11 from the bottom) of 56 mm thick iron
plates with 45 mm gap between each iron layer where 10 layers of RPCs,
the active detector elements, are placed. Each iron layer of mini-ICAL
is made up of 7 plates of iron named as A, B, C and D according to
their dimensions (given in Table~\ref{tab:plate_dim}) and positions
(shown in Fig.~\ref{final_mini_ical_up}). The total area span of each
layer is $4 \times 4$ m$^2$.  In three layers, viz., 1 (bottom), 6
(middle) and 11 (top), small gaps of 3--4 mm have been introduced
to enable insertion of Hall sensor PCB of thickness $\sim$ 2.3 mm
to measure the magnetic field. These gaps, numbered 0--5 are shown in
Fig.~\ref{final_mini_ical_up}; gaps 1, 4 are designed to be 4 mm thick and
980 mm long, while the remaining gaps are designed to be 3 mm in width
and 800 mm in length. In the remaining layers these gaps are about 2 mm
in width. The diagonal line shown in Fig.~\ref{final_mini_ical_up}
is a reference line along which magnetic field values are extracted from
the simulated model, as we will discuss later. As can be seen from the
geometry, the magnetic fields are expected to be similar across the gaps
0, 2, 3, and 5, and similar across gaps 1 and 4.

A set of two OFHC copper (Ref: procured from Luvata, Austria) coils, each 
having 18 turns and cross section area 30 $\times$ 30 mm$^2$ is used to 
magnetize the mini-ICAL detector. The copper conductor used to make the 
coils has a central bore of diameter 17 mm through which is circulated 
low conductivity ($<$ 10 micro siemens per cm) water for cooling purposes.
The total resistance of the coils is about 7.6 milli-ohm. The power supply 
used to magnetize the detector is a high current (50 Volt DC and 1250 Amp 
DC) with high stability of $\sim$ 100 ppm (that is within 90 milli-amps while 
supplying 900 amps to the magnet). The power supply is designed by National 
Power Research Laboratory, Indore. With the current on in the coils (i.e. 
when the magnet is on) along with chilled water supply, the outside surface 
temperature of the copper coil reaches 20-20.5 $^0$C. It takes only a 
couple of minutes to stabilize to this value. This temperature is much 
lower than the room/ambient temperature, which is maintained at about 
28 $^0$C degrees for the optimum operation of RPCs.

\begin{table}[htp]
\centering 
\begin{tabular}{|c|c|c|} \hline  
Type of iron plate  & Size                     & No. of plates   \\ \hline 
A                   & 2001 $\times$ 1000 mm    & 2               \\ \hline
B                   & 2001 $\times$ 1000 mm    & 2               \\ \hline
C                   & 1200 $\times$ 2000 mm    & 2               \\ \hline
D                   & 1962 $\times$ 1600 mm    & 1               \\ \hline
\end{tabular}
\caption{Each iron layer of mini-ICAL consists of 4 types of iron plates
as listed here and shown schematically in Fig.~\ref{final_mini_ical_up}.}
\label{tab:plate_dim}
\end{table}

\section{Magnetic field measurements in the mini-ICAL}

\subsection{The Hall sensor probe and calibration}

The procedure followed for the calibration of the Hall sensor PCB and
estimation of error associated with the measured magnetic field is the
same as described in Ref.~\cite{Khindri:2022elz}, where measurements
were presented for coil currents upto 500 A due to limitation of power
supply. In the present paper, measurements are shown for currents upto
900 A, where saturation of magnetic field is expected.

\begin{figure}[htp]
  \centering
    \includegraphics[width=0.8\linewidth, height=0.13\linewidth]{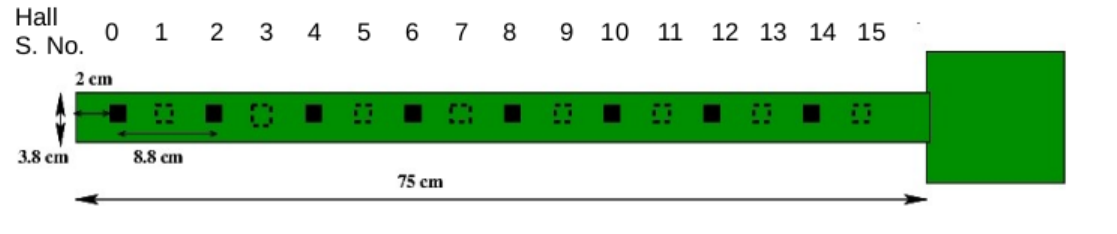}
  \caption{Hall probe PCB with 16 Hall sensors mounted on it}
  \label{fig:Hall_PCB}
\end{figure}

Magnetic field measurements are done in the gaps~0--5 shown in
Fig.~\ref{final_mini_ical_up} at coil current of 500 to 900 A in
steps of 100 A using Hall probe PCB whose schematic is shown in
Fig.~\ref{fig:Hall_PCB}. The probe has 16 CYSJ106C GaAs \cite{data:sheet} 
Hall Effect Element sensors mounted on it and these are labeled from 0 
to 15. Closer to the front end electronics is sensor no.~15 and the farther 
one is numbered as sensor no.~0. The probe is 75 cm in length and 3.8 cm in
height. The Hall sensors are mounted 4.4 cm apart from each other and on
alternate sides of it such that sensors with even numbers are on one side
and sensors having odd sensor numbers are on the other side of the PCB.
The thickness of Hall probe PCB becomes after mounting the sensors is 2.3
mm.

\begin{figure}[thp]
  \centering
    \includegraphics[width=0.49\textwidth, , height=0.35\textwidth]{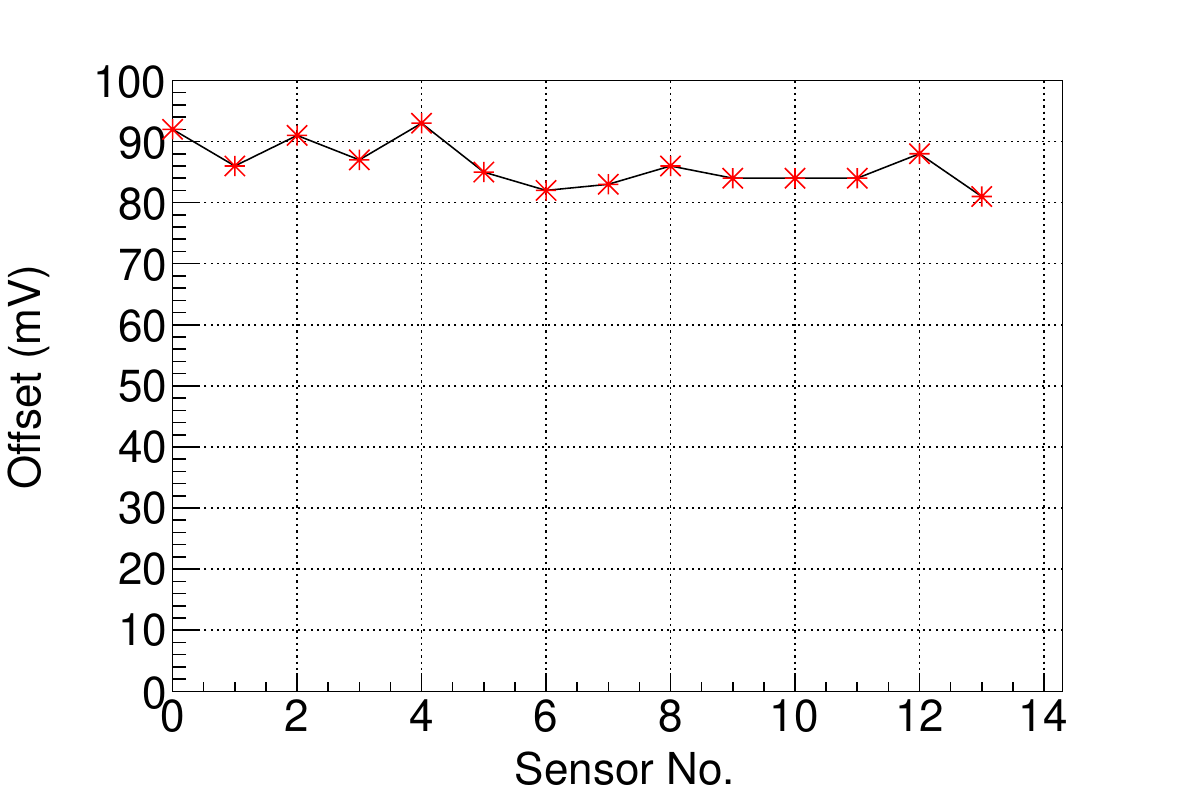}
    \includegraphics[width=0.49\textwidth]{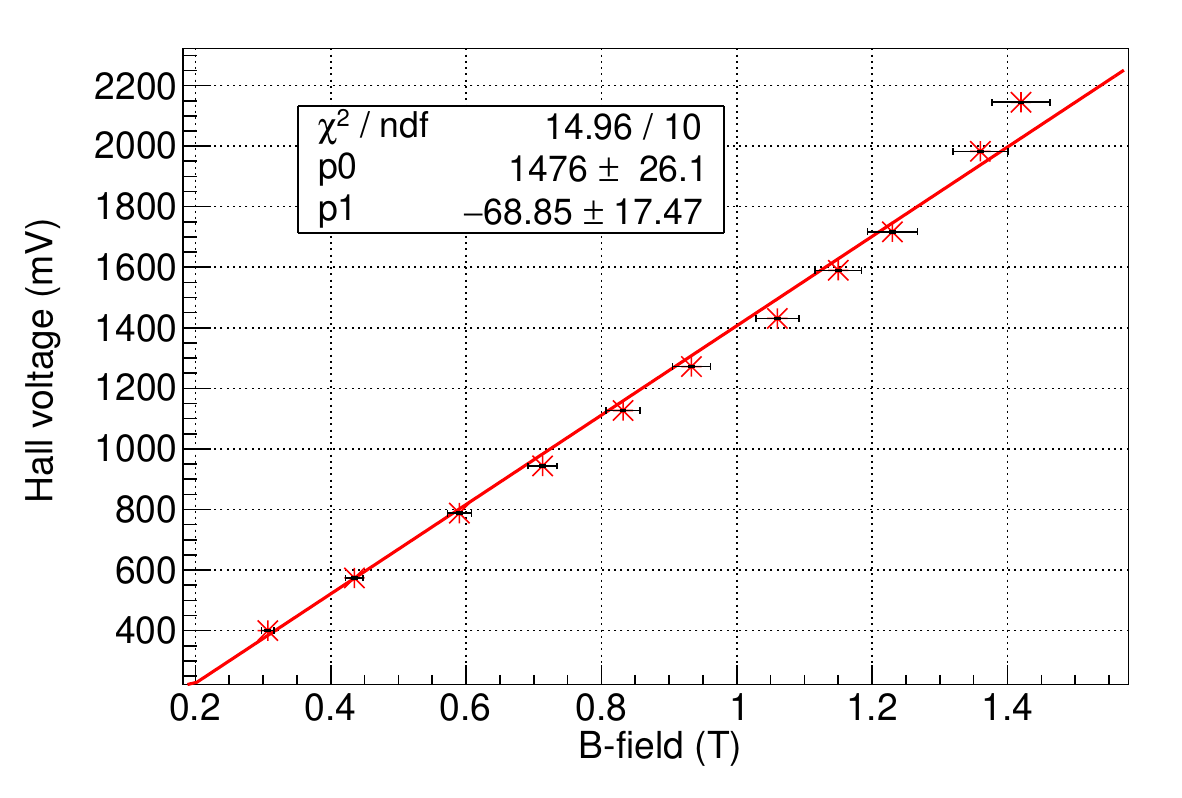}
    \caption{Left: Measurement of offset for all the sensors; 
    Right: Calibration plot of sensor number~8.}
    \label{calibration_plot}
\end{figure}

Before taking measurements of the field at mini-ICAL, the calibration and
offset measurement of each sensor is done. A detailed description of the
procedure can be found in Ref.~\cite{Khindri:2022elz}; the same procedure
was used for the current set of measurements as well. To measure the
offset, the Hall sensor is kept away from the mini-ICAL (any) magnetic
field and the output voltage of each sensor is measured and plotted
(see the left side of Fig.~\ref{calibration_plot}). The calibration
of each sensor is done using an electromagnet designed and fabricated for 
the purpose of calibration. The field in the gap has been cross-calibrated 
using a precision Hall probe. The output Hall voltage of each sensor is 
noted for different magnetic field values. This is plotted and fitted with 
linear fit to extract the slope and intercept. A sample calibration curve 
for sensor number 8 is shown in the right side of Fig.-\ref{calibration_plot}.

\begin{figure}[htp]
  \centering
    \includegraphics[width=0.55\linewidth, height=0.36\linewidth]{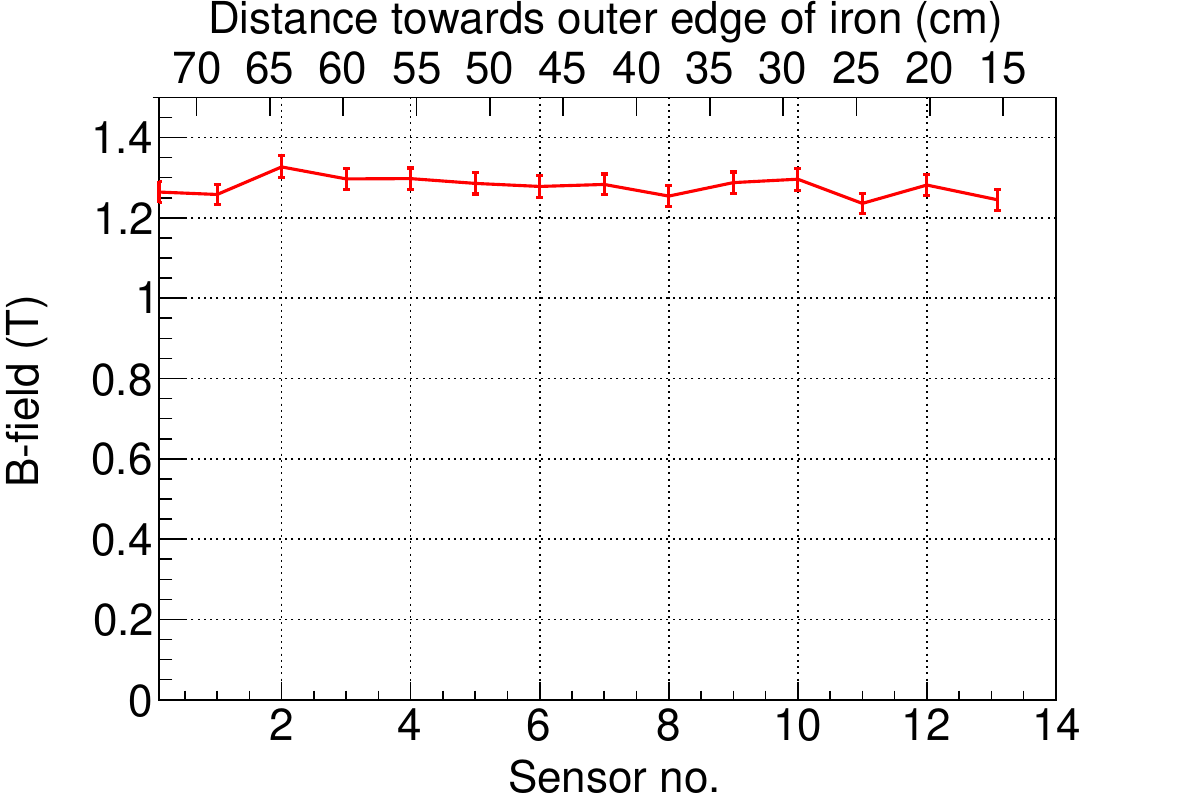}
  \caption{Sample magnetic field measurement in the gap-2 at 900 A coil current
  as obtained from the calibration and offsets discussed in the text}.
  \label{900_amp_gap_2}
\end{figure}

The magnetic field is calculated from
\begin{eqnarray} \nonumber
B  & = & \frac{V-V_{0}+V_{\epsilon}}{m}~, \nonumber \\
\delta {B} & = & \frac{B}{m} \sqrt{\frac{\Delta {V_{0}}^2}{B^2} +\Delta
m^2}~,
\label{eq:formula}
\end{eqnarray}
where $V$ is the measured Hall voltage, $V_{0}$ is the measured offset
voltage (for each sensor), $m$ is the slope obtained from the calibration
curve, and a small stray offset of V$_{\epsilon} \lesssim 20$ mV was
included to get agreement between the odd and even sets of sensors. Here
the errors in the measurement is $\delta {B}$, which depends on the
error in offset measurement, $\Delta {V_{0}}$ and the error in the slope
from the calibration curve $\Delta m$.  In Fig.~\ref{900_amp_gap_2} the
magnetic field measurement\footnote{Since sensors 14 and 15 were defective, 
results are shown for sensors 0 to 13.} in the gap-2 is shown for the 
coil current of 900 A. Here the sensor number~0 is nearest to the 
coil and sensor number~15 is nearest to the outer edge of the iron layer. 
The upper $x$-axis of Fig.~\ref{900_amp_gap_2} is labelled according to 
the actual distance of each sensor from the edge of the detector/layer.

\subsection{Magnetic field measurements in the top layer}

\begin{figure}[htp]
  \centering
    \includegraphics[width=0.49\textwidth]{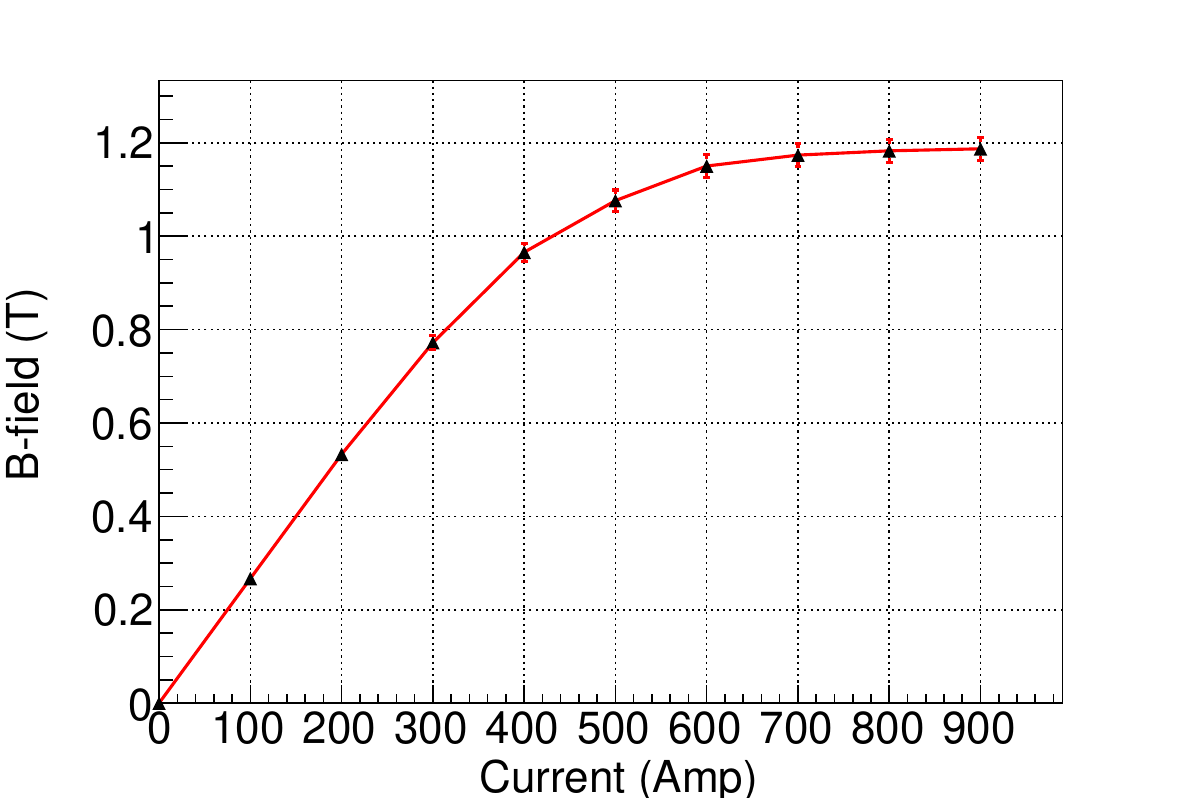}
    \includegraphics[width=0.49\textwidth]{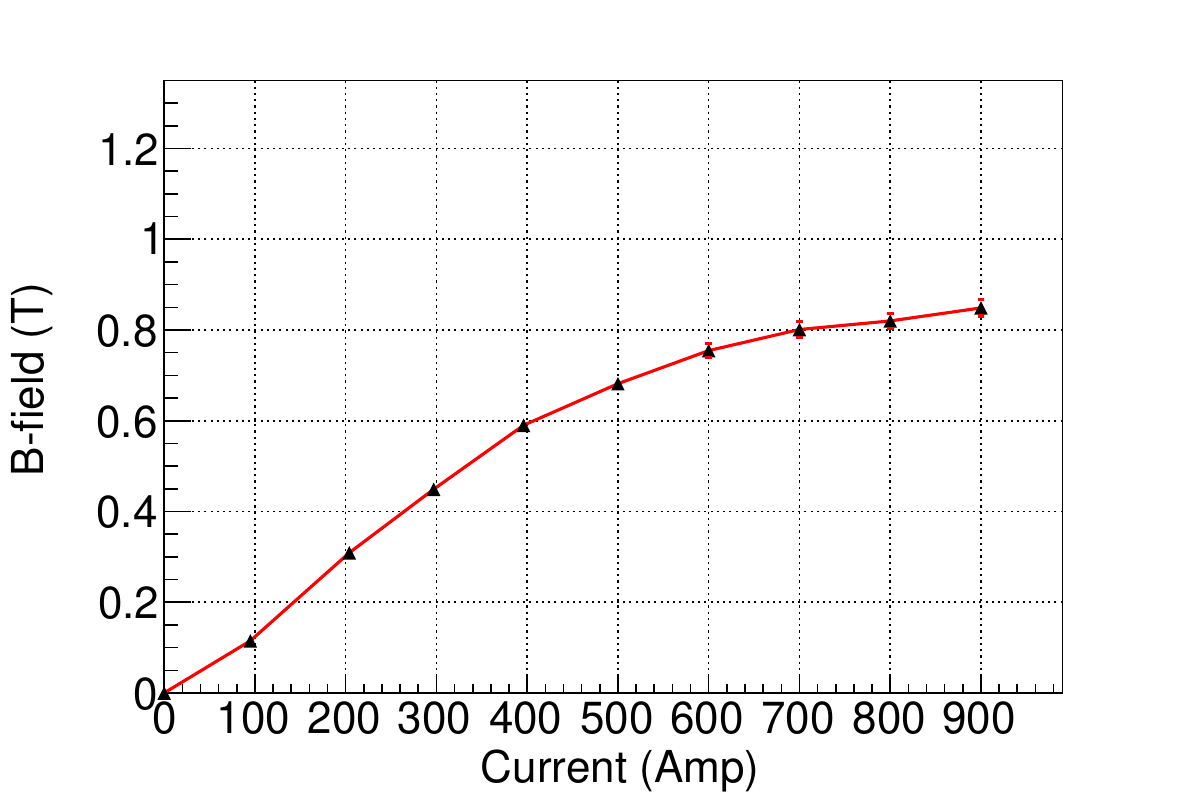}
\caption{Measured magnetic field in the gaps 0 (left) and 4 (right) as a
function of coil current in steps of 100 A. } 
\label{fig:B_vs_I}
\end{figure}

The measured magnetic field is plotted as a function of the coil current,
in steps of 100 A up to a maximum of 900 A, in Fig.~\ref{fig:B_vs_I}
for gaps 0 and 4. The results for gaps 2,3,5, are similar to that for
gap 0, and gaps 1 and 4 have similar behavior.  It can be seen that in
the gap-0 (length 800 mm), the magnetic field reaches saturation at 900
A while the magnetic field in the gap-4 (length 980 mm) is still in the
linear region and approaching saturation. Also, the magnitude of the
field is smaller for the horizontal gap 4 than for the vertical gap 0.
This behavior is also seen in the simulations study, as we will see later.

\begin{figure}[htp]
  \centering
    \includegraphics[width=0.49\textwidth]{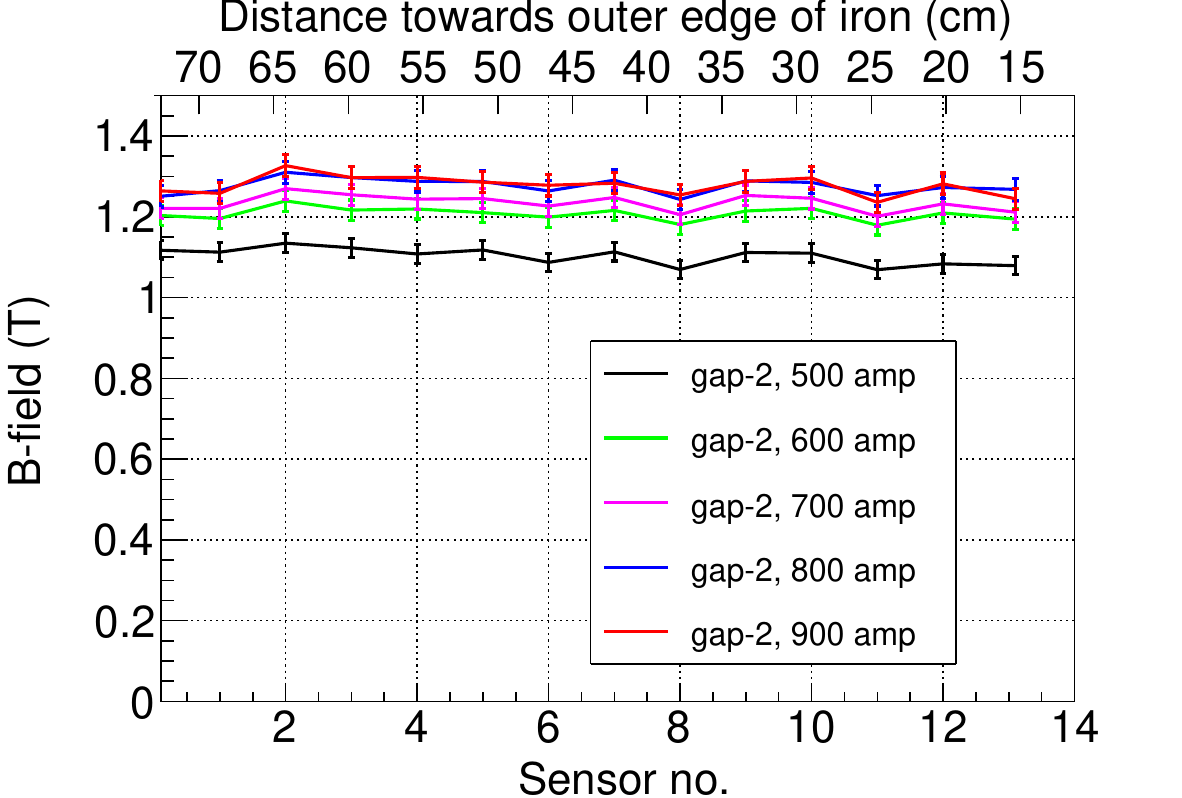}
    \includegraphics[width=0.49\textwidth]{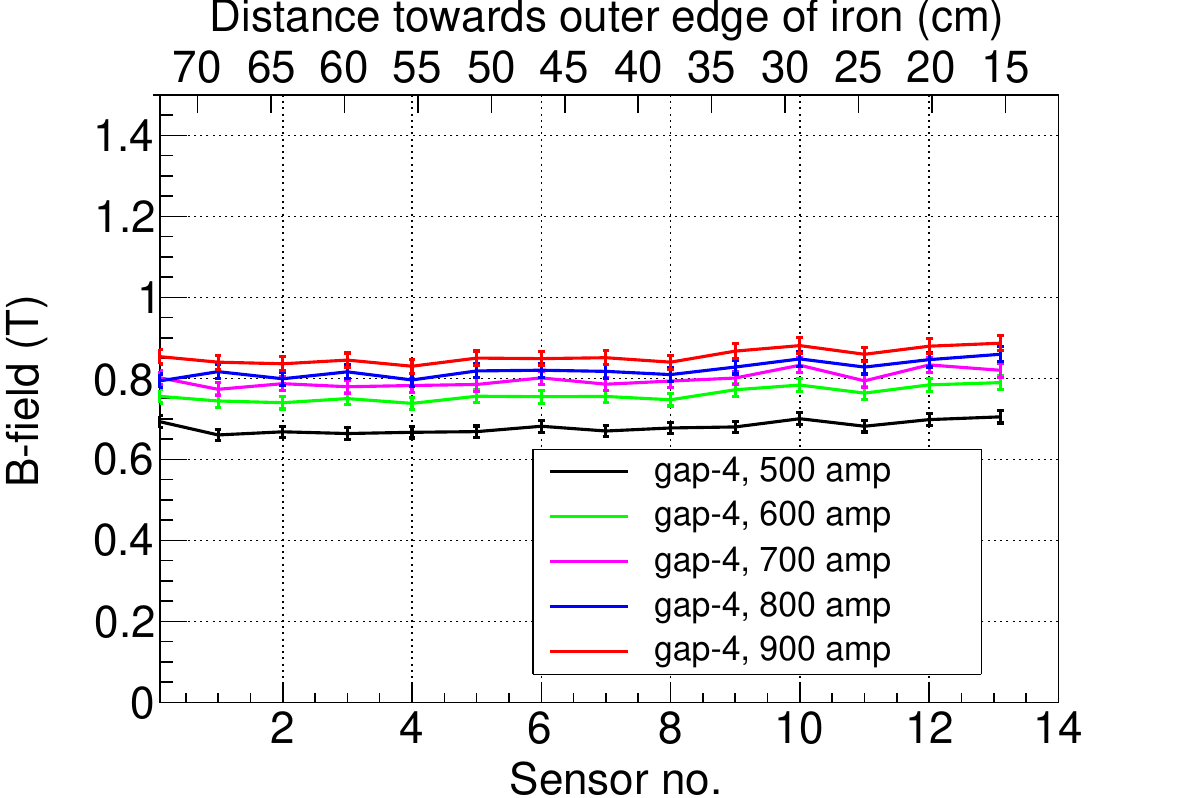}
    \caption{Measured magnetic field in the gaps 2 (left) and 4 (right)
    for coil currents from 500 to 900 A in steps of 100 A as a function
    of the distance from the detector edge.}
\label{fig:B_distance}
\end{figure}

The magnetic field as measured by the Hall probe sensors is plotted for
the sample vertical gap-0 in Fig.~\ref{fig:B_distance} as a function
of the distance from the detector edge, for coil currents, 500, 600,
700, 800 and 900 A. It can be seen that the field is more or less
uniform across the gap, and increases with current, saturating just as
shown in Fig.~\ref{fig:B_vs_I}. A similar result is seen for the
horizontal gap-4 as well, as also seen in Fig.~\ref{fig:B_distance}. A
small increase in the field is seen (for all values of coil current)
towards the coil end of the gap. 

\begin{figure}[htp]
  \centering
    \includegraphics[width=0.49\textwidth]{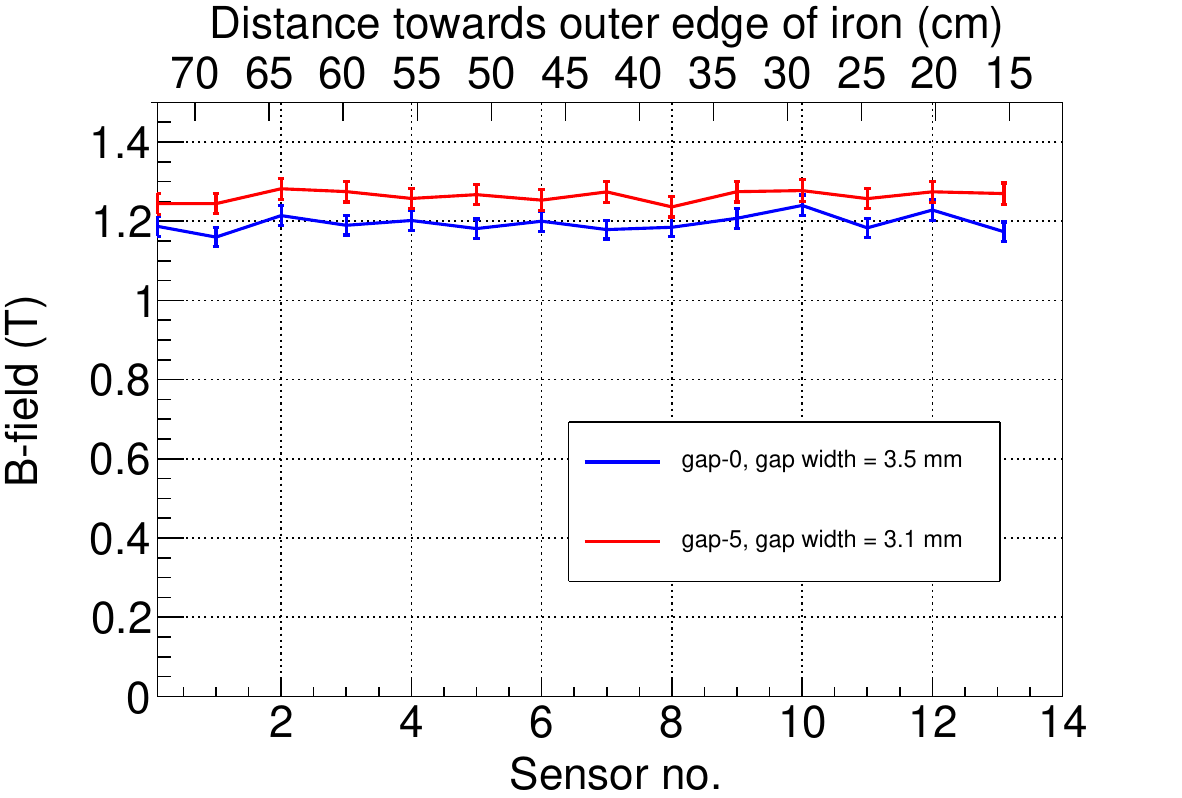}
    \includegraphics[width=0.49\textwidth]{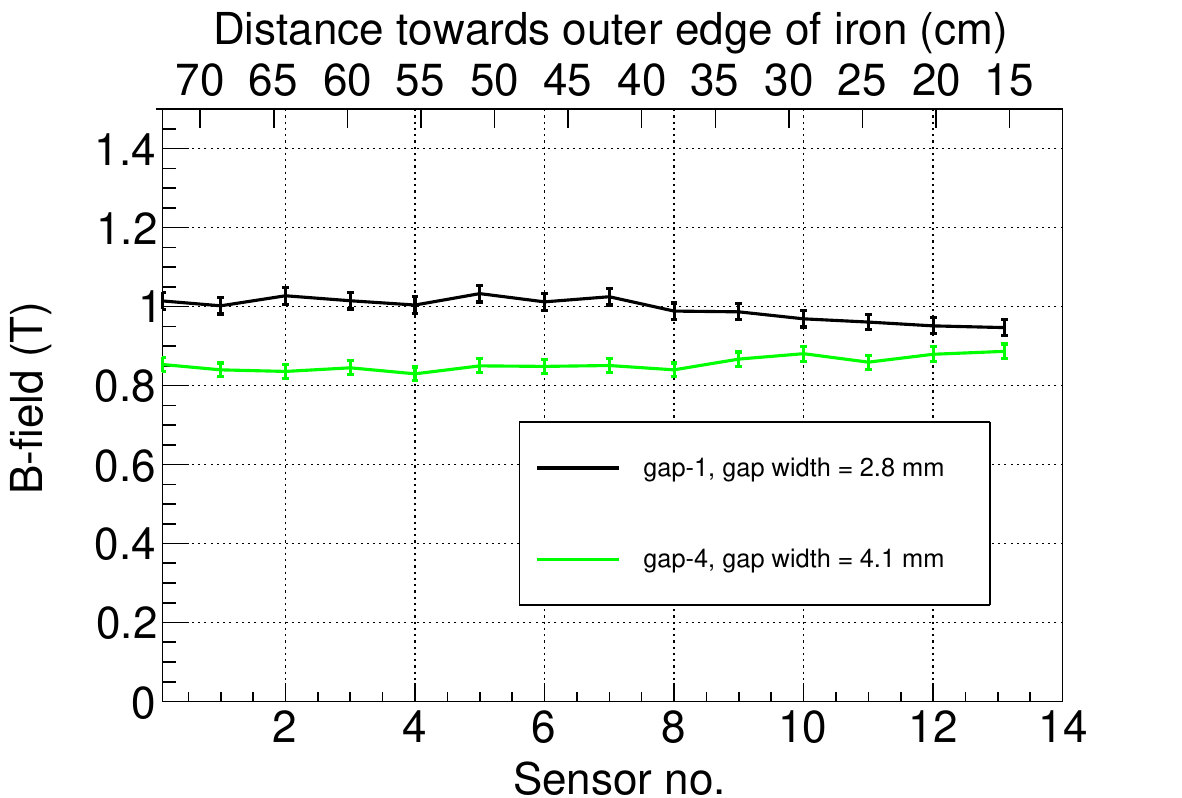}
\caption{Measured magnetic field in the gaps 0,5 (left) and 1,4 
        (right) for coil current of 900 A as a function of the distance from 
        the detector edge. These are expected to show similar behaviour as per 
        the detector geometry. See text for more details.}
\label{fig:B_distance_2}
\end{figure}

Fig.~\ref{fig:B_distance_2} shows the same results, but for two different
vertical gaps, 2 and 3, for a sample coil current of 900 A, which are 
expected to show similar behaviour. The RHS of the figure shows the same, 
but for horizontal gaps, 1 and 4 (gap-width shown in the Fig.~\ref{fig:B_distance_2} 
is averaged value, however gap - 1 is a bit wider near the edge and hence 
magnetic field near edge is less than the magnetic near the coil as can 
be seen Table.~\ref{tab:gap_width_new} and near the outer edge width of gap-4 
is less than width near the coil therefore magnetic field is a bit more than 
magnetic field near the coil as can be seen in Table.~\ref{tab:gap_width_new}). 
The difference in the magnitudes of the fields at these gaps is found to be 
\cite{Khindri:2022elz} due to variations of the actual gaps from the design 
values. This has already been discussed for the measurement at 500 A in our 
earlier study and is visible for other values of coil current as well. 
Hence it is important to adjust for the gap widths. So, before going on 
with the description of the magnetic field measurements in mini-ICAL, we
discuss details of the various gap widths.

\subsection{Gap width measurements}
\label{sec:gap_width}

It is observed that the width of the gaps which are introduced in the
mini-ICAL to measure the magnetic field is different from the designed
width. Gaps 0, 2, 3 and 5 were designed with width of 3 mm while gaps 1
and 4 were designed with widths of 4 mm. These gaps widths were slightly
different from the design values due to assembly tolerances. In addition,
it was found that the gap widths changed perceptibly when the coil current
was turned on, due to torsional forces from the magnetic field. However,
beyond a current of 500 A, there was no appreciable change found in the 
width as the current increased up to 900 A. These gap widths affect the 
magnetic field measured in the gaps as already discussed in 
Ref~.\cite{Khindri:2022elz}, with gaps having smaller width showing higher 
magnetic field and vice versa. Hence an accurate measurement of the various 
gap widths is important in order to deduce the correct magnetic field.


These gap widths for gaps in the top layer of mini-ICAL were
measured, on the top side, using (non-magnetic) brass vernier calipers 
and the values are shown\footnote{Access to gaps in other layers, or even 
on the bottom side of the top layer, is very difficult due to the narrow 
space between the iron plates (even when the RPCs are removed).} in 
Table~\ref{tab:gap_width_new}).  Furthermore, it is observed that these 
gaps are varying in width, across their lengths, as can be seen from 
Table~\ref{tab:gap_width_new}. It was shown in our earlier work that 
adjusting for the different gap widths results in matching of the 
magnetic field across similar gaps, {\em viz.}, across gaps 0,2,3,5, 
and gaps 1,4. We will discuss this further once we have presented 
detailed studies of the magnetic field simulations.

\begin{table}[htp]
\centering 
\begin{tabular}{|c|c|c|c|c|c|c|} \hline  
Gap number $\to$       &   0     & 1    & 2     & 3    & 4     & 5  \\ \hline
Measurements taken $\downarrow$ & \multicolumn{6}{|c|}{(in mm)} \\ \hline
near coil              & 3.5    & 2.8   & 3.2  & 3.1  & 4.1  & 3.1   \\ \hline
at middle              & 3.5    & 2.8   & 3.4  & 3.4  & 4.2  & 3.1   \\ \hline
at outer edge of iron  & 3.5    & 2.9   & 3.2  & 3.6  & 3.9  & 3.0   \\ \hline
\hline
Average value used     & 3.5    & 2.8   & 3.2  & 3.4  & 4.1  & 3.1   \\ \hline
\end{tabular}
\caption{Measured gap widths of the gaps 0--5 in the top layer of
mini-ICAL, showing both the range of values as measured along their
widths and their average, when the current is switched on. The measured
widths are the same for currents in the range 500--900 A.}
\label{tab:gap_width_new}
\end{table}

\section{Magnetic field simulation using Infolytica MagNet software}

\begin{figure}[htp]
 \centering
\includegraphics[width=0.65\textwidth]{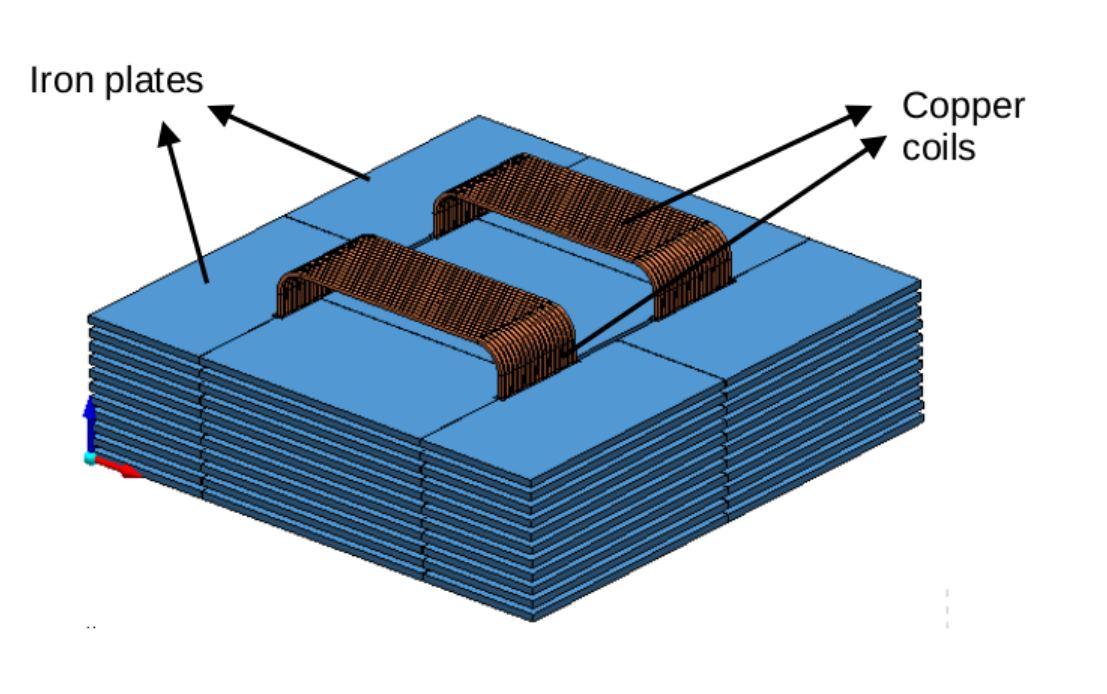}
\caption{Schematic of the mini-ICAL geometry encoded in MagNet software.}
\label{fig:mini_ICAL_geom}
\end{figure}

A 3-D static magnetic field simulation is done using the Infolytica MagNet
version 7.7 software \cite{magnet6} and the mini-ICAL geometry as one of the 
inputs. It uses finite element methods to calculate the magnetic field in
the magnet volume. It divides the given geometry into small volume elements
and solves the Maxwell's equations in that part using approximate
methods. The mini-ICAL geometry coded into the MagNet software is shown
in Fig.~\ref{fig:mini_ICAL_geom}. In what follows, an ideal mini-ICAL
geometry is considered. After assembly of the mini-ICAL some changes in
the dimensions are observed, such as variations in the actual gaps as
described in Section \ref{sec:gap_width}, particularly after switching
on the coil current, and their effect is discussed later.

We are interested in simulating the magnetic field over the entire
mini-ICAL as well as, in particular, the $B$-field in the gaps. Since the
mini-ICAL geometry is similar to ICAL, we shall display the simulated
field along the diagonal ($x = y$ in mini-ICAL.  As can be seen from
Fig.~\ref{final_mini_ical_up}, this line nicely captures the variation
in $B$ from $B = 0$ at $x = y = 0$ mm at the bottom-left corner through
the maximum value at the center, $x = y = 2002$ mm to the top-right
corner, $x=y=4004$ mm. It also traverses two 3 mm gaps between the C
and D plates, which are regions of interest.

\subsection{Inputs to the simulation}

There are certain inputs that are required to perform the simulations. One
important input is the $B$-$H$ curve which defines the behavior of the
iron with respect to magnetization. Chemical composition of the steel 
used for the mini-ICAL is given in the Table~\ref{tab:iron_comp}, low carbon 
steel (from Essar Steel, Hazira, Gujarat) is used to build mini-ICAL and 
this shows a saturation magnetic field of 2.099 T; see Table~\ref{tab:iron_char}. 
The other inputs for the simulations are the plate and coil geometry, 
mesh size and shape, and coil current.

\begin{table}[htp]
\centering 
\begin{tabular}{|c|c|c|c|c|c|c|c|c|} \hline  
Elements & C     & Mn & Si & P & S & N & Al & Fe          \\ \hline 
\%       & 0.015 & 0.368 & 0.188 & 0.012 & 0.008 & 0.005 & 0.001 & Balance\\ \hline
\end{tabular}
\caption{Chemical composition of the mini-ICAL steel. }
\label{tab:iron_comp}
\end{table}

\begin{table}[htp]
\centering 
\begin{tabular}{|c|c|} \hline  
Material           & Low carbon steel          \\ \hline 
Young's modulus    & 200 GPa         \\ \hline
Density            & 7850 kg/m$^3 $       \\ \hline
Poisson's ratio    & 0.3        \\ \hline
Yield strength     & 200 MPa (Min.)      \\ \hline
Magnetic property  & Knee point 1.5 T, H1T 300 A/m, H1.5T 900A/m \\ \hline
\end{tabular}
\caption{Mechanical and magnetic properties of the steel used for
mini-ICAL. Source: Essar Steel.}
\label{tab:iron_char}
\end{table}

\subsubsection{$B$-$H$ curve}

\begin{figure}[htp]
 \centering
\includegraphics[width=0.49\textwidth]{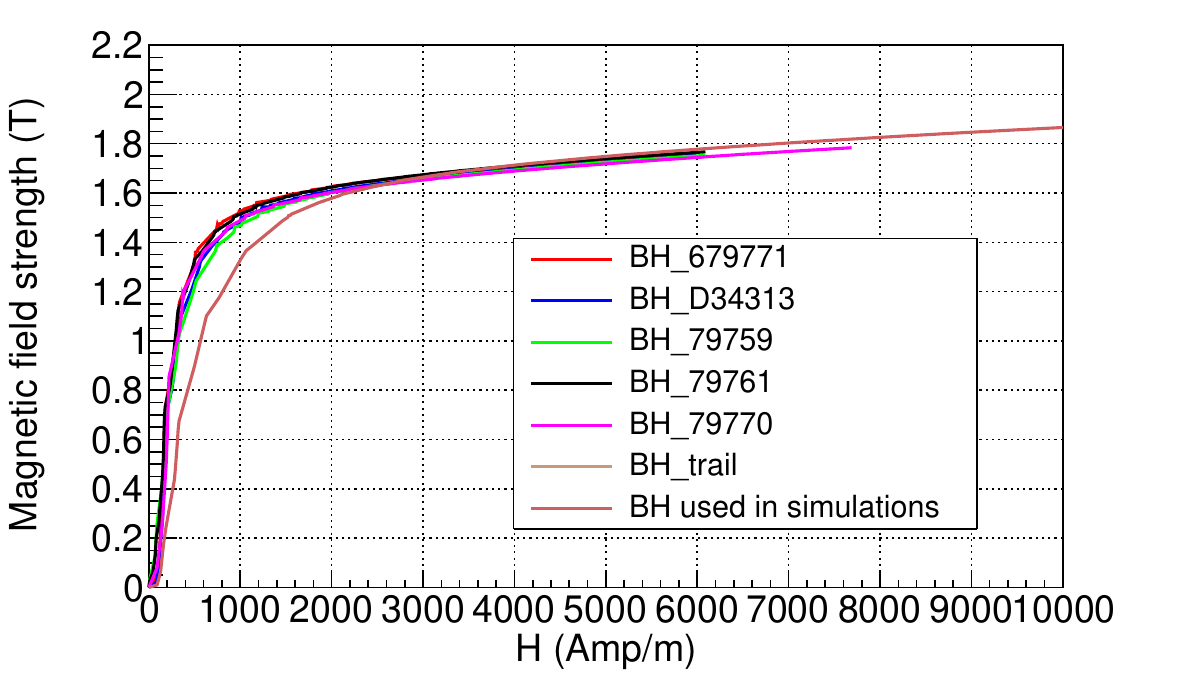}
\includegraphics[width=0.49\textwidth]{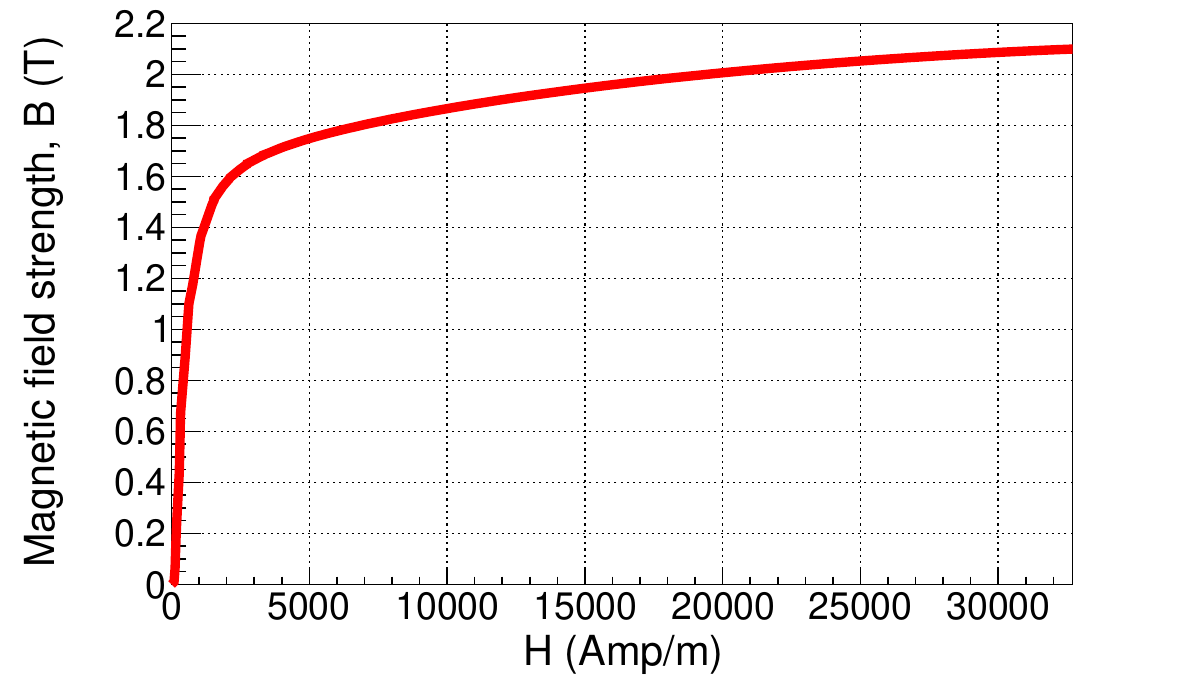}
\caption{Left: Various measurements of the $B$-$H$ curve of
different samples using a calibrated gaussmeter. Right:
The $B$-$H$ curve used as an input in magnetic field
simulations.}
\label{fig:B_H_curve}
\end{figure}

The $B$-$H$ curve is a basic property of the material. It is used as
an input parameter for the simulations and therefore it is important
to measure it as accurately as possible. The $B$-$H$ curve is measured
(with a commercial $B$-$H$ tracer) using a ring specimen with square
cross section made from the same low carbon steel used in mini-ICAL.
The Gaussmeter used in the measurement
has an error of 1\% in the 1 T magnetic field range, and is calibrated
against a reference magnet \cite{gaussmeter}.
The results of various data sets match with the reference input
$B$-$H$ curve, shown in Fig.~\ref{fig:B_H_curve}, which was used as an input
for the MagNet 7.7 simulations.

\subsubsection{Mesh size and shape}

It has been observed that mesh size plays a crucial role in the simulations.
Simulations done using crude or default mesh (of size $\sim$ 300--400 mm) 
shows magnetic field values different than the simulations done using finer 
mesh (size $\sim$ 1--2 mm), especially where the field is changing
rapidly. However it is observed that for even finer mesh 
size there is no significant change in magnetic field values while taking 
a much larger computation time and memory. Therefore optimization of the input 
mesh sizes is important.

Different mesh sizes are optimized for iron plates and gaps since iron 
plates ($\sim$ 1 m -- 2 m) are of larger dimensions and gaps (dimensions 
$\sim$ 800 mm $\times$ 3 mm) are of small dimensions as compared to the 
iron plates. The mesh size for mini-ICAL geometry is optimized based on 
three factors---memory available in the system, simulation time and 
convergence in the magnetic field value. Since gap widths are smaller 
(3--4 mm), the mesh size used here was 2 mm. For the iron plates with 
larger dimensions (1000 mm--1200 mm), a mesh size of 10 mm was used.

\begin{figure}[htp]
  \centering
    \includegraphics[width=0.49\textwidth]{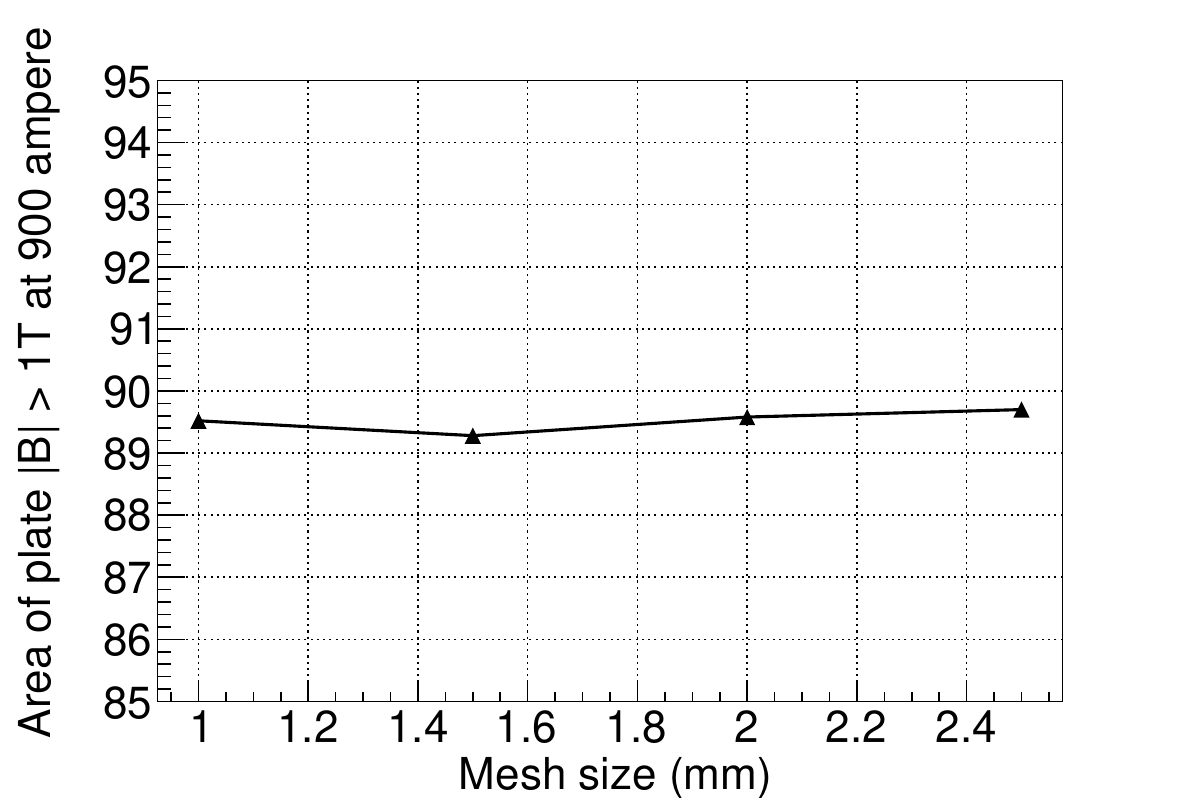}
    \includegraphics[width=0.49\textwidth]{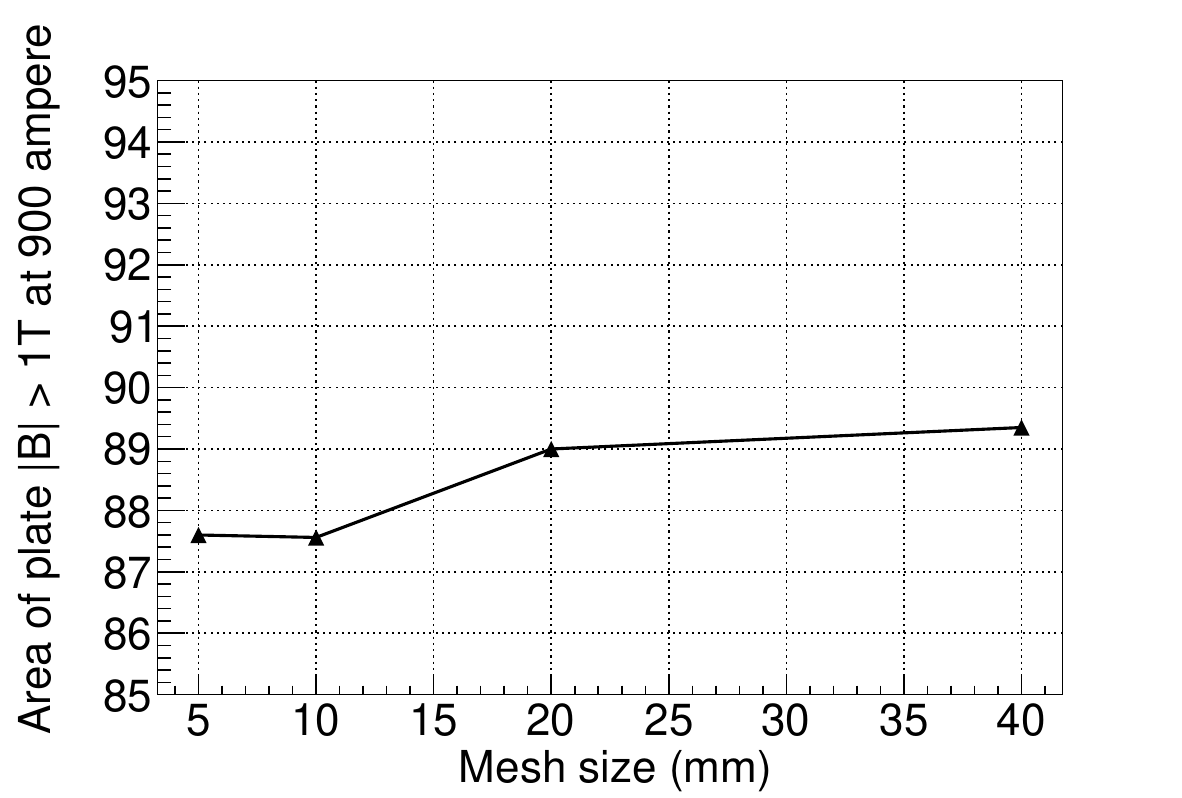}
\caption{Fractional change in volume having $|B| > 1$ T of a layer with
change in mesh size in gaps (left) and iron plates (right).}
    \label{fig:mesh_size}
\end{figure}

To optimize the mesh size the magnitude of the field, $|B|$, is extracted
for the complete layer and the fraction of area (\%) having $|B| > 1$
T is calculated with 900 A coil current and the change in the fraction
is noted. It can be seen in Fig.~\ref{fig:mesh_size} there is no significant
change in the \% fraction of area with magnetic field value $> 1$ T below
2 mm of mesh size in the gap. Since the computation time for simulating
the complete model is an acceptable 6--7 hours (on a desktop computer) 
for this mesh size in the gap, this was used in all subsequent simulations.

A similar procedure was adopted to optimize
the mesh size in the iron plates. From Fig.~\ref{fig:mesh_size}, it
can be seen that the fraction of area (\%) having $|B| > 1$ T does not
change for a mesh size below 10 mm while increasing the computation
time and memory for smaller mesh sizes. Hence a mesh size of 10 mm was
used for the iron plates in all simulations. Another metric for deciding 
the mesh size in the gap could be the convergence of the B-field there.

\begin{figure}[!tbp]
  \centering
    \includegraphics[width=0.49\textwidth]{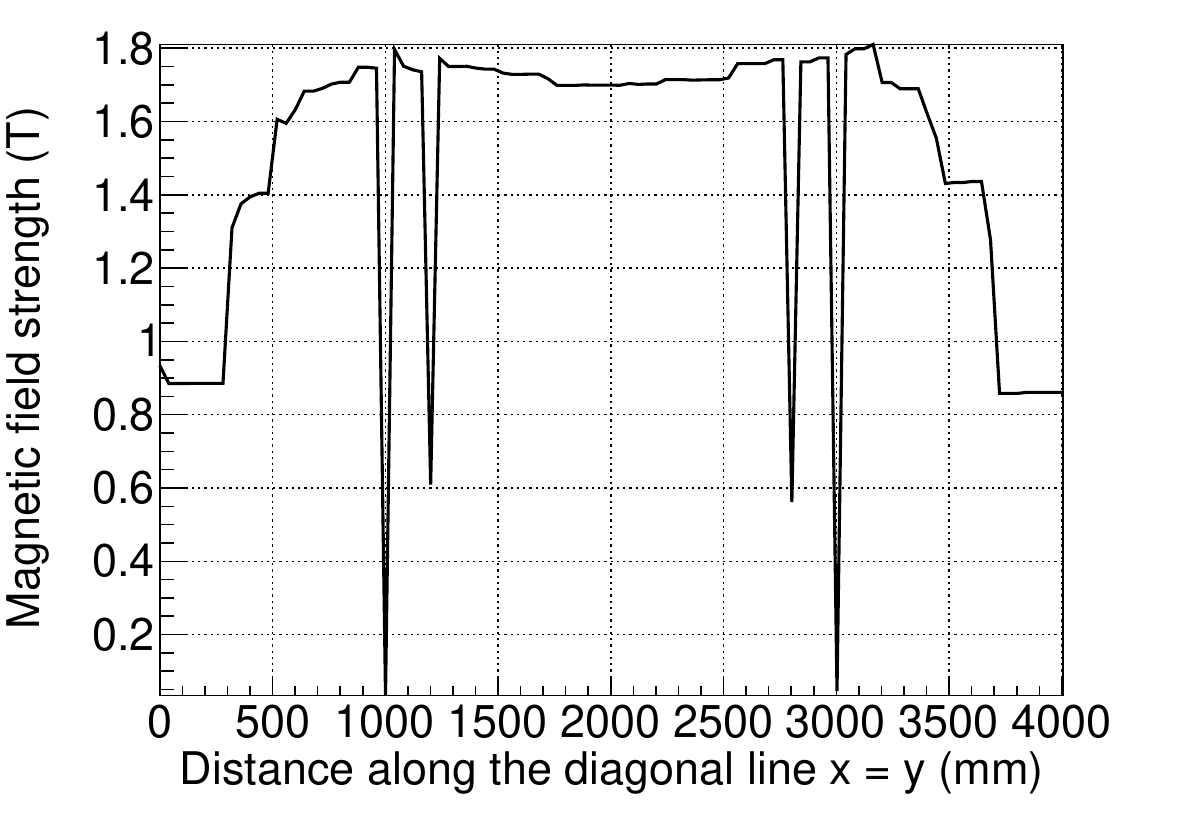}
    \includegraphics[width=0.49\textwidth,height=0.35\textwidth]{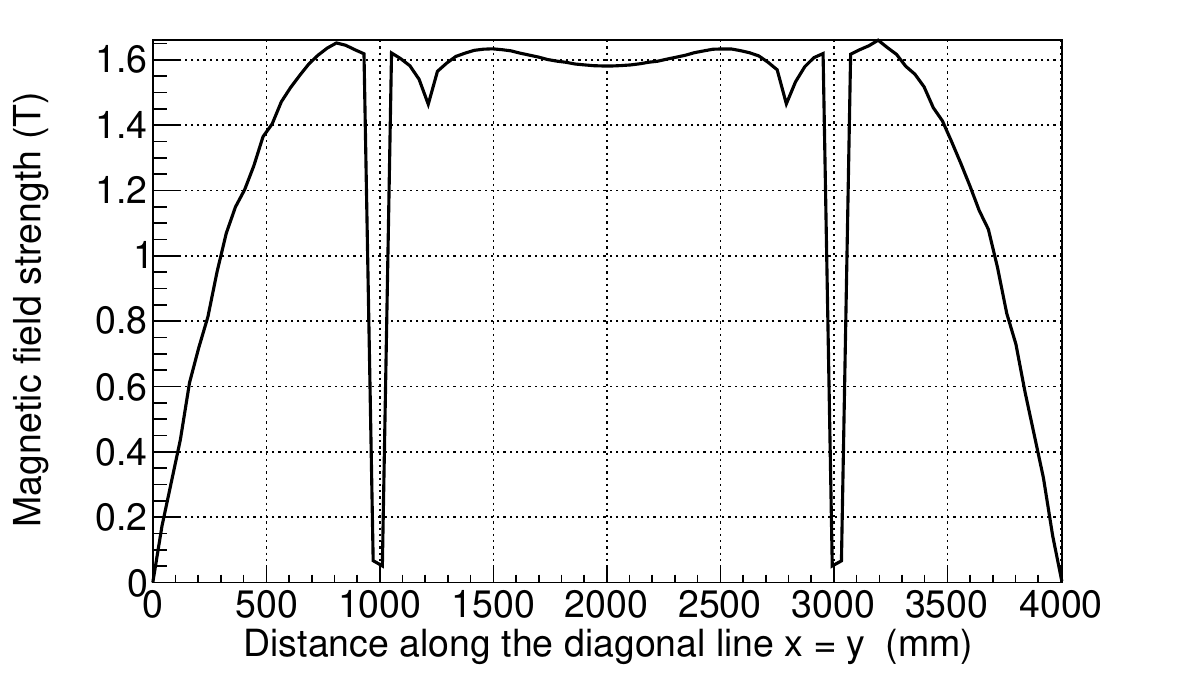}
    \caption{Magnetic field along the diagonal line of the top layer for 900 
    A coil current simulated with default mesh size $\sim 300$--400
    mm (left) and with mesh size of 2 mm in the gaps and 10 mm in the
    iron plate (right). Note the $y$-axes are different in the two
    plots.}
    \label{with_without_mesh}
\end{figure}

The change in magnetic field value with default and optimized mesh size
is visible in Fig.~\ref{with_without_mesh}. It is observed that the
magnetic field value extracted along the diagonal line ($x=y$) of the
mini-ICAL layer for the model simulated with optimized mesh (2 mm in
the gap and 10 mm in the iron) is more smooth and symmetric as compared
to the $B$-field values extracted from the model simulated with default
mesh size (300--400 mm). In addition, the predictions for the $B$-field
in the gap between C and D layers (at $x=y=1200$ mm and $x=y=2800$ mm)
are very different between the two sets of simulations. Since we are going
to validate the simulations by comparing the values of the $B$-field in
the gap regions (in gaps 0--5) to the measured values, it is crucial to
precisely determine the values in these gaps.

Also the model simulated with default mesh size shows magnetic field
value which is larger than the model simulated with optimized mesh
size. Therefore it can be seen that the optimization of mesh size is one
of the important aspects of simulations.  It may be mentioned that the
shape of the mesh element used in the simulation has a tetrahedral shape.

\subsubsection{Coil Current}

There are two sets of copper coils in mini-ICAL, each having 18 turns, and DC 
current is passed through these coils to magnetize the mini-ICAL magnet. The
mini-ICAL magnetic field is simulated for 500, 600, 700, 800, 900 A of coil 
current. Since 500 A lies in the linear region of B-H curve it will give an 
insight into the magnetic field behavior in the linear region of B-H curve. 
Therefore simulations are performed for coil currents between 500 A and 900 A 
(near saturation of B-field) in steps of 100 A. Also, the results for 500 A 
current will be compared with the measurements made in the mini-ICAL detector 
currently functioning at Madurai.

\subsection{Comparison between single layer and 11-layer model}
\label{sec:no_of_layers}

\begin{figure}[htp]
 \centering
\includegraphics[width=0.65\textwidth]{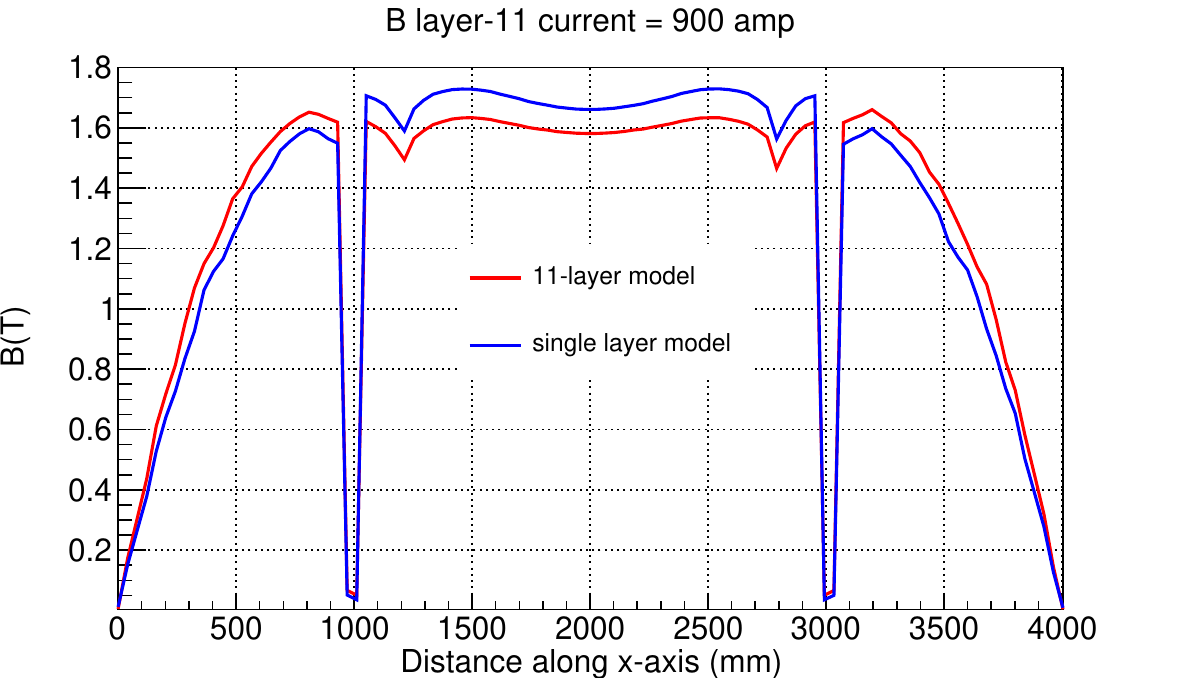}
\caption{Comparison of magnetic field values along the diagonal line
(see Fig.~\ref{final_mini_ical_up}) for is plotted to show the
11-layer and single layer model.}
\label{fig:11_vs_single_lay}
\end{figure}

Simulations are performed to study the change in magnetic field value for
models of mini-ICAL having all the 11 layers and a single iron layer. This
is done to find out if the $B$-field value does not change
significantly; as a single layer or 3-layer model will require much
less time and memory for the simulation. In Fig.~\ref{fig:11_vs_single_lay}
the $B$-field value along the diagonal line (see
Fig.~\ref{final_mini_ical_up}) is plotted to show the
difference and it can be seen that in the middle plate i.e., D (central
region between the two major dips) the $B$-field value is about 6\%
more for the model simulated with single iron layer and in the outer
plates, the magnetic field value is about 3\% more for the model simulated
with 11 iron layers.

From here it can be concluded that there is a significant difference
between the magnetic field value calculated using a single layer and
11-layer model, although the origin of this difference is a bit subtle.
It has to do with the counteracting effect of the field in the plates
above and below the layer. Hence a 11-layer model was used for the
simulation study although it requires more time and memory. A more
detailed analysis of the variation in magnetic field due to the number
and position of iron plates is given in Appendix \ref{app:a}.

\subsection{Simulated magnetic field in mini-ICAL}

\begin{figure}[bhp]
 \centering
\includegraphics[width=0.60\textwidth]{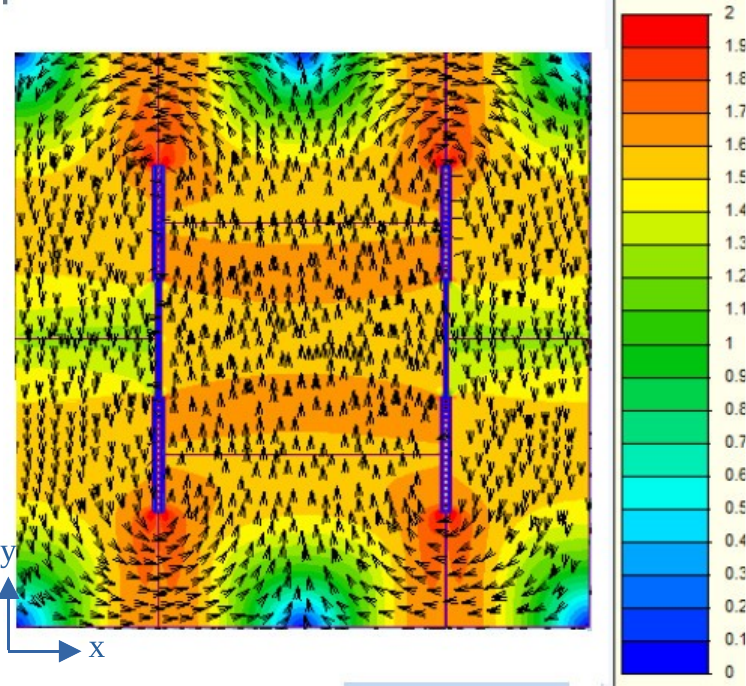}
\caption{Magnetic field map simulated for the top layer of mini-ICAL
for 900 A coil current.}
\label{fig:B_map}
\end{figure}

After simulations, a magnetic field map is generated for the layer
of mini-ICAL (as generated for ICAL detector). The magnetic
field map generated for one layer of mini-ICAL can be seen in
Fig.~\ref{fig:B_map}.

This shows a close resemblance to the $B$-field map for the ICAL detector
\cite{sp:bhr}. The $B$-field is maximum and uni-directional (in the $y$
direction) in the central region; in the side region (outside the coil
slots, but in the central $y$ region from approximately 1000--3000 mm), it
is in the opposite direction and a little less than in the central region
and in the peripheral region (outside 1000--3000 mm in both $x$ and $y$
directions), it is varying in direction as well as magnitude. This also
agrees with the measured magnetic field, where the field in the vertical
slots (0,2,3,5) is much higher than in the horizontal slots 1,4, as seen
from Fig.~\ref{fig:B_distance}.

\begin{figure}[htp]
 \centering
\includegraphics[width=0.85\textwidth]{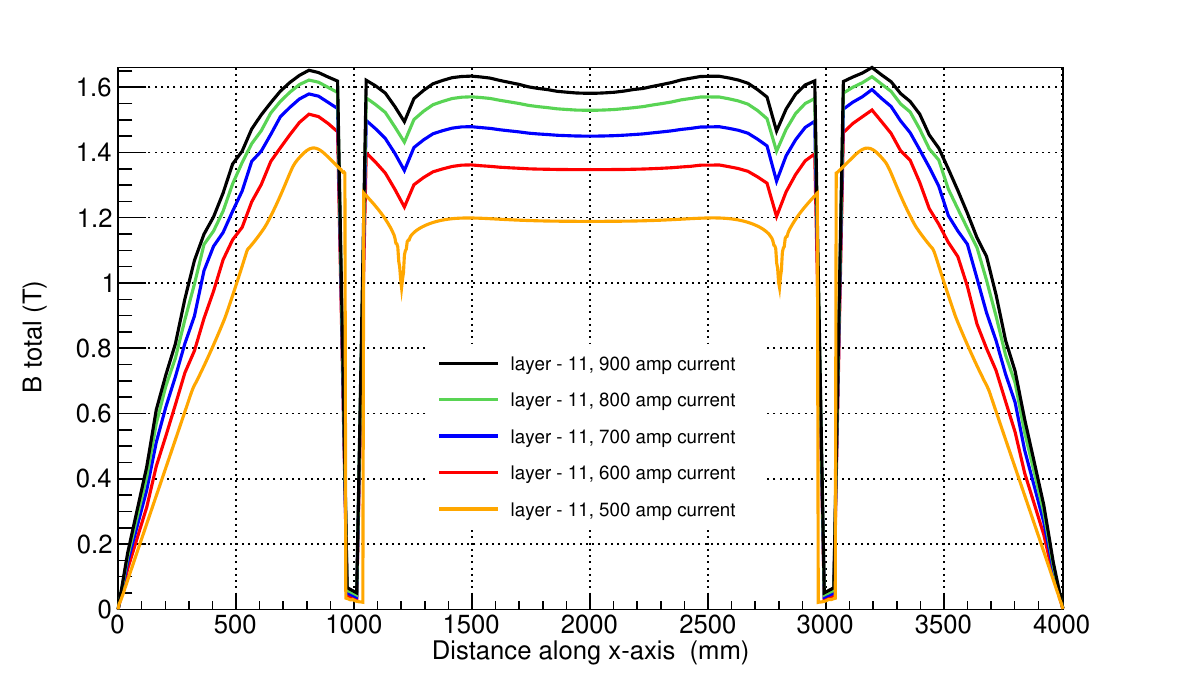}
\caption{Magnetic field along the diagonal line (see
Fig.~\ref{final_mini_ical_up}) for the top layer of
mini-ICAL at 500 A, 600 A, 700 A, 800 A and 900 A coil currents.}
\label{B_diag}
\end{figure}

Data is also extracted for the magnetic field along the diagonal line
(see Fig.~\ref{final_mini_ical_up}) and in the gaps 0--5. The $B$-field
value along the diagonal is shown in Fig.~\ref{B_diag} for different
values of the coil current. It can be seen that as the current approaches
900 A from 500 A the variation in the magnetic field values reduces due
to the onset of saturation. It can be seen from Fig.~\ref{final_mini_ical_up}
that the diagonal line crosses two coil slots and two gaps between the
iron plates and hence gives a fair idea of the $B$-field over
the entire detector. The four dips in the magnetic field along the
diagonal line---two larger dips corresponding to the two coil slots
(width - 80 mm) and two smaller ones corresponding to the gap between
the iron plates C,D and D,C (which are approximately 3 mm in width)
are thus explained.

\begin{figure}[htp]
  \centering
    \includegraphics[width=0.49\textwidth, height=0.35\textwidth]{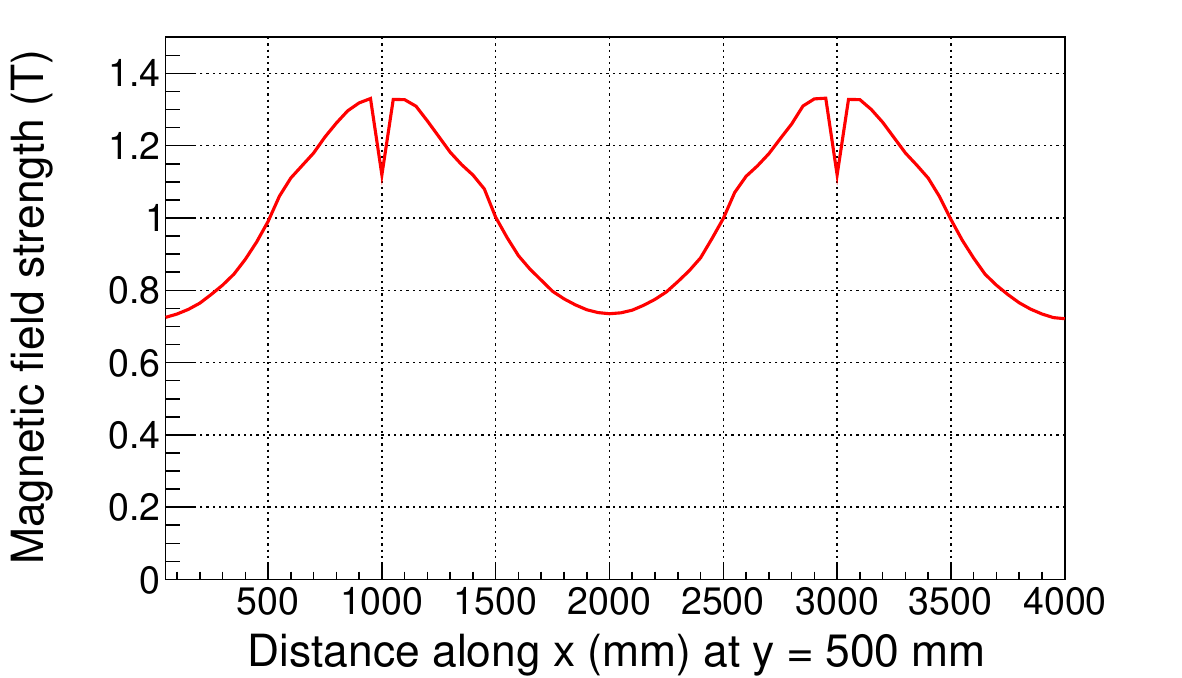}
    \includegraphics[width=0.49\textwidth, height=0.35\textwidth]{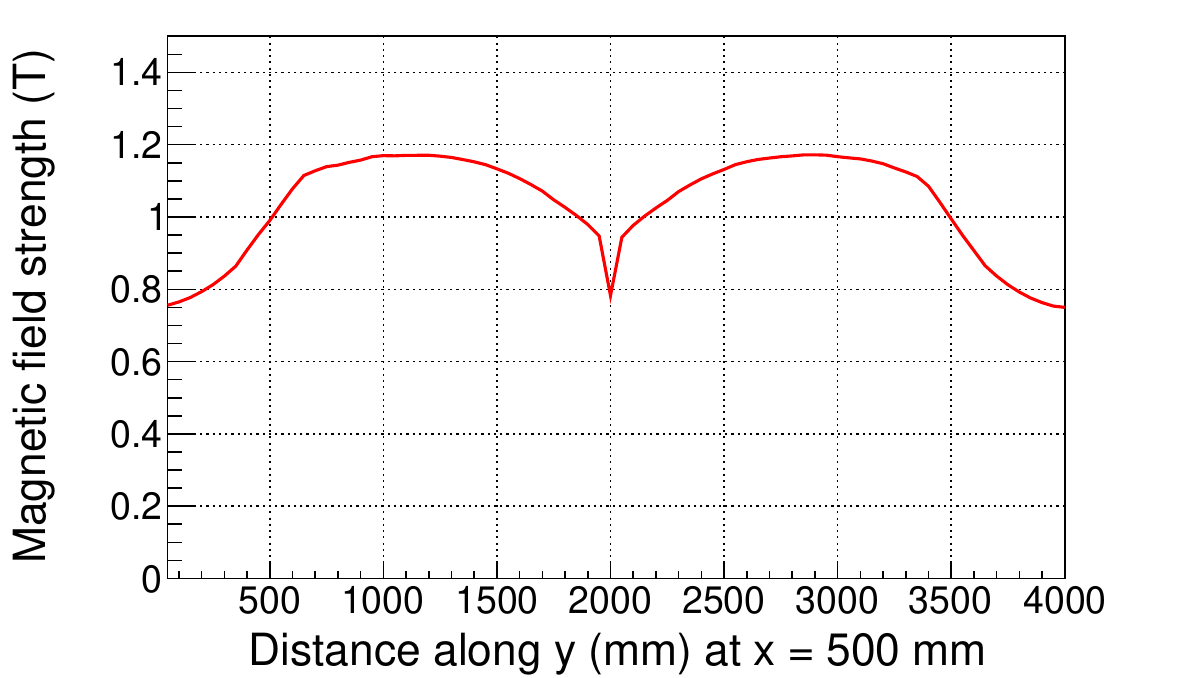}
\caption{Left: Simulated magnetic field $\vert B \vert$ for (Left:) $y = 500$ mm and for $x =
0$--4004 mm, including gaps 2,3 and (Right:) $x =
500$ mm and along $y = 0$--4004 mm, including gap 1.}
    \label{sim_B_along_x_y_500mm}
\end{figure}

Fig.~\ref{sim_B_along_x_y_500mm} (left) shows the magnetic field value
along the line $y = 500$ mm for $x = 0$--4004 mm and the two dips are due
to the gaps-2 and 3 (width - 3mm).  In Fig.~\ref{sim_B_along_x_y_500mm}
(right) the simulated magnetic field is shown along the line $x = 500$ mm,
for $y = 0$--4004 mm and the dip in the magnetic field is due to presence
of air gap-1 (width - 4mm) as shown in Fig.~\ref{final_mini_ical_up}. These
are the gaps at which the magnetic field has been measured at mini-ICAL
using Hall probes. Note that in the simulations, the field is symmetric
with respect to gaps (0,3), (2,5) and (1,4).

\section{A detailed study of the simulated $B$-field in the gaps}

In order to make a meaningful comparison between the measured and
simulated magnetic fields, we need to determine the magnetic field in
the gaps where the measurements have been made.
The magnetic field values are extracted for all of the gaps where the
magnetic field measurements at mini-ICAL have been done, to compare
the simulated $B$-field with the measured field. It was already shown
in Ref.~\cite{Khindri:2022elz} that although all the gaps --- 0, 2,
3 and 5 are symmetric and ideally are of same width but the $\vert
B\vert$ field values measured in these gaps are slightly different from
each other which was due to variation in gap width. To study this, and
account for these differences, simulations are performed by varying the
gap widths.

\subsection{Dependence of magnetic field on gap width}

\begin{figure}[htp]
  \centering
    \includegraphics[width=0.49\textwidth]{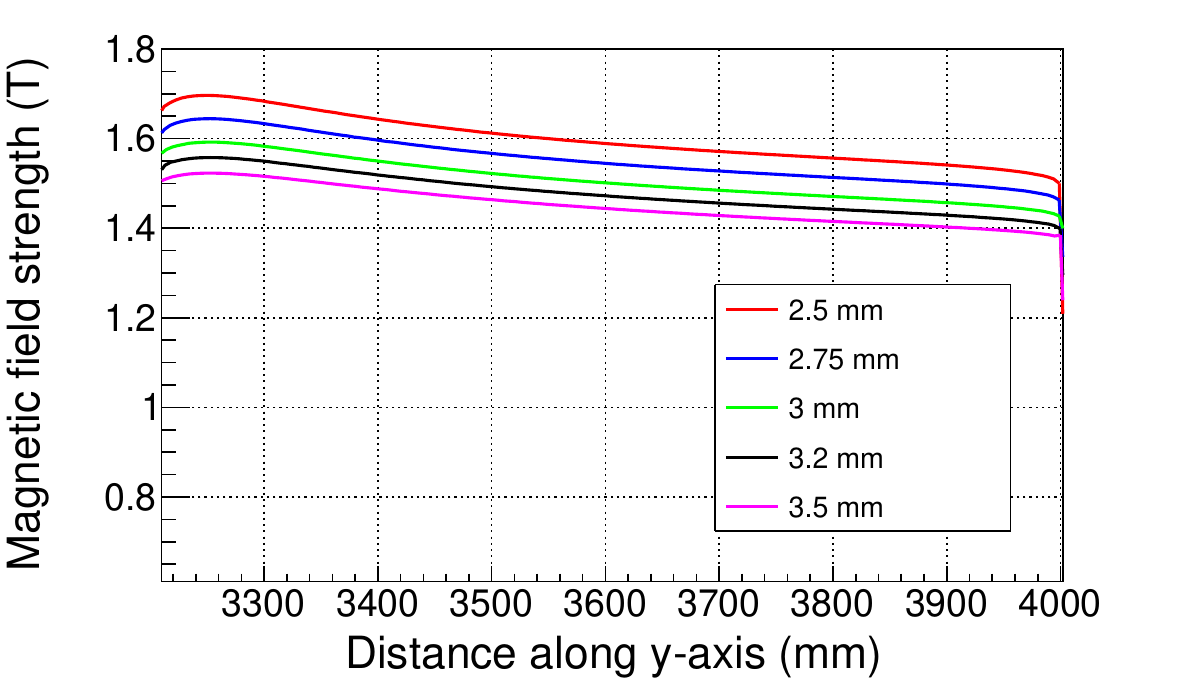}
    \includegraphics[width=0.49\textwidth]{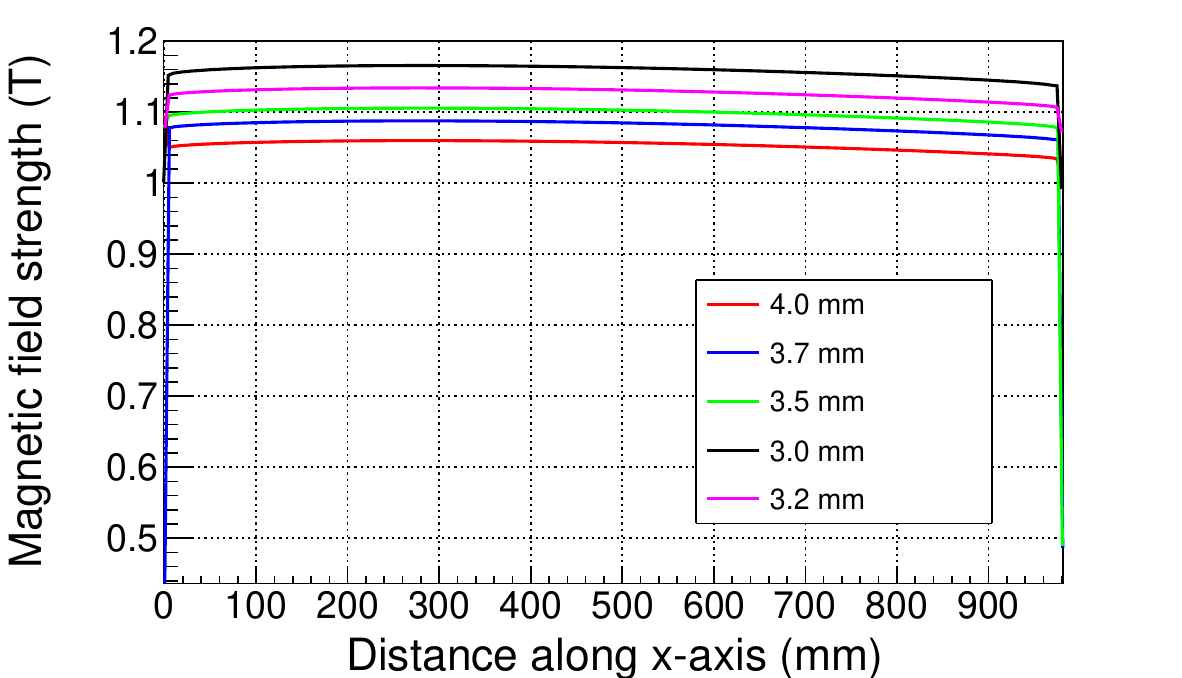}
\caption{Simulated magnetic field values in gap 0 (left) and gap 1
(right) for varying gap widths; the larger distance values correspond to
points closer to the detector edge.}
\label{gap_width_01}
\end{figure}

Since gaps 0, 2, 3, and 5 are identical by design (width 3 mm), and gaps
1 and 4 are also designed (width 4 mm) to be identical to each other, due
to the symmetry of the geometry, we study the variation in the $B$-field
due to changes in gap width for the representative gaps 0 and 1 only. The
gap width for gap 0 is varied from 2.5 to 3.5 mm in steps of 0.25 mm and
for gap 1 it is varied from 3 to 4 mm in steps of 0.25 mm and $B$-field is
studied at these gaps. These are shown in Fig.~\ref{gap_width_01} for
gaps 0 (left figure) and 1 (right figure). On the left side, the
magnetic field value is extracted for a fixed $x$ value corresponding to
the position of the gap 0, with $y$ varying from $y=3200$--4000 mm, from
inside the detector, to the outside edge. For gap 1, the $y$-value
corresponds to the position of the gap 1, and the distance varies from
$x = 0$--980 mm from a point just
outside the coil to the detector edge. Note that the gaps lengths are
800 mm for gap 0 and 980 mm for gap 1, since the Hall probe is 660 mm in 
length, it probes practically the entire length of the gaps 0, 2, 3, and 
5 (barring 140 mm near the coil) while it cannot probe 320 mm also of the 
gaps 1 and 4 near the coil slots. Hence comparison between simulations and 
data is done, keeping this in mind. While the $B$-field is roughly constant 
over gap 1, it falls off with distance in gap 0.

\begin{figure}[h]
 \centering
\includegraphics[width=0.49\textwidth]{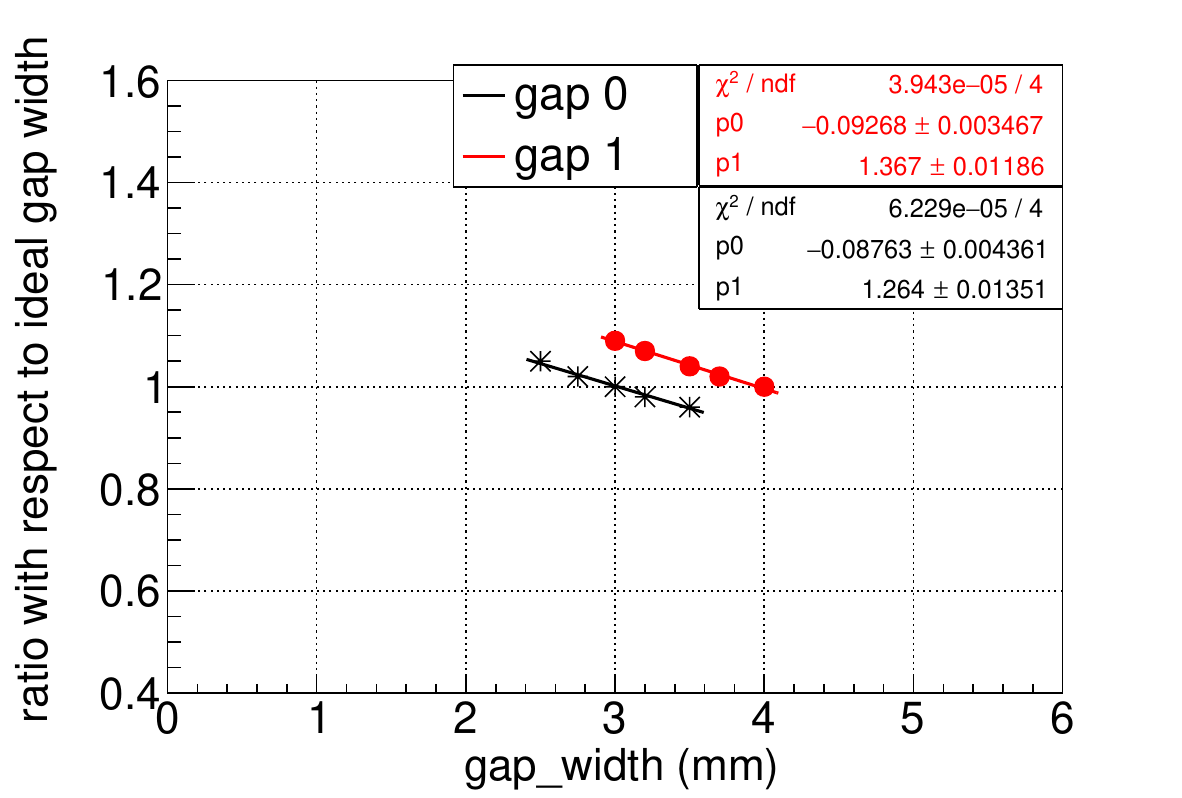}
\caption{Scaling of magnetic field as a function of the gap width.
Plotted here is the ratio of the $B$-field values obtained with a given
gap width (as given along the $x$-axis) to the value of the $B$-field
expected when simulating with the design gap width of 3 mm for gap 0 and
4 mm for gap 1. Also shown is a linear fit to the ratio, for use by
interpolation to determine the simulated field at arbitrary gap values.}
\label{gap_para}
\end{figure}

Fig.~\ref{gap_para} shows the change of $B$-field obtained due to
variations in the gap width. Shown are the ratios of the results
obtained by simulating the field for a given gap at an arbitrary
position along the gap, to that obtained for a simulated gap with the
design width (of 3 mm for gap 0 and 4 mm for gap 1). The results are
robust and independent of the position chosen. From the $B$-field values 
extracted for the simulations with different gap widths, it is observed 
that the dependence on gap width can be parameterized by a linear fit. 
This can be used to calculate the $B$-field for any gap width by interpolating 
in the region of interest. It was also seen that either of the two fits,
linear and parabolic, does not make much difference in the region of
interest. Hence we shall assume a linear dependence on the gap width. The
ratio corresponding to the actually measure gap width, as listed in
Table~\ref{tab:gap_width_new}, was found through interpolation of the data
and fits determined in Fig.~\ref{gap_para}. This was used to determine the
simulated $B$-field values corresponding to the actual gap values found
in the measurement, by adjusting for the gap width.

\section{Comparison of simulated and measured  magnetic fields}

Recall that the gaps 0, 2, 3, and 5 all have a design width of 3 mm 
while gaps 1 and 4 have a design width of 4 mm; furthermore, as per the
detector geometry, the gaps 0,2,3,5 are expected to show similar
behaviour, and gaps 1,4 are expected to be similar; hence we consider 
these two sets separately.

We are now ready to compare the results of the simulation (for different
coil currents) with the measurements of the $B$-field in the gaps
discussed in the earlier section. In each case, we plot
the ratio of the simulated to measured magnetic field with and without
gap width corrections.

\begin{figure}[htp]
  \centering
    \includegraphics[width=0.49\textwidth]{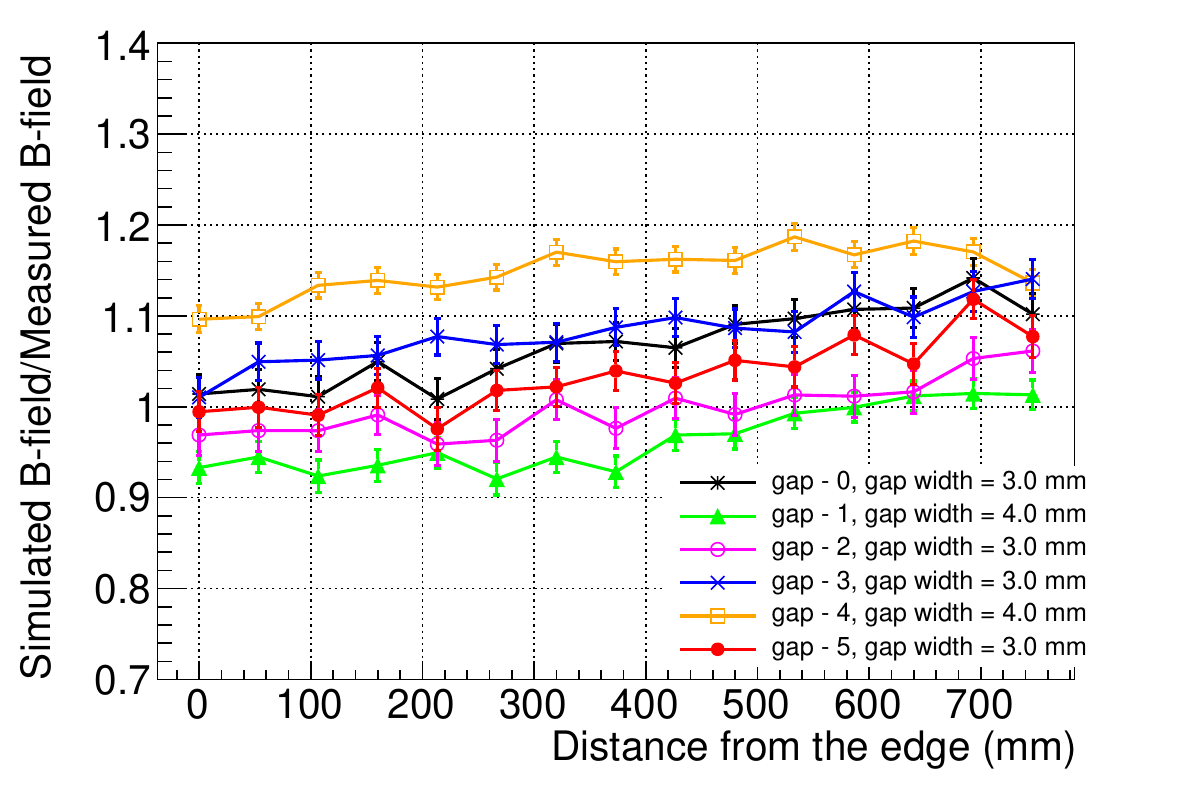}
    \includegraphics[width=0.49\textwidth]{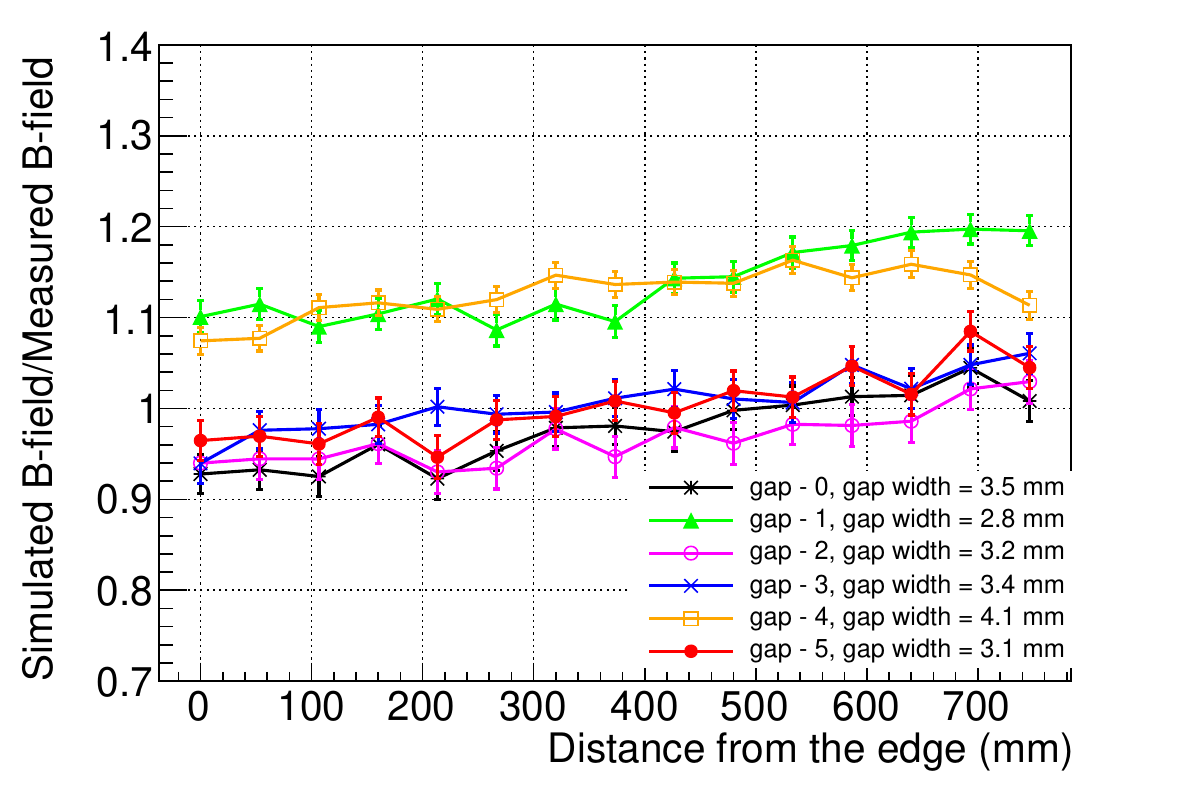}
\caption{Left: Comparison of measured magnetic field with simulations done
using design values of gap widths at the coil current of 500 A; Right:
Comparison of measured magnetic field with simulations done with corrected
gap widths (as per Table~\ref{tab:gap_width_new}) at the coil current of
500 A. The agreement between various measured magnetic fields has improved
but the ratios for the gaps-(1,4) and (0,2,3,5) are bunched together.}
\label{fig:sim_vs_meas_500_amps}
\end{figure}

\begin{figure}[htp]
  \centering
    \includegraphics[width=0.49\textwidth]{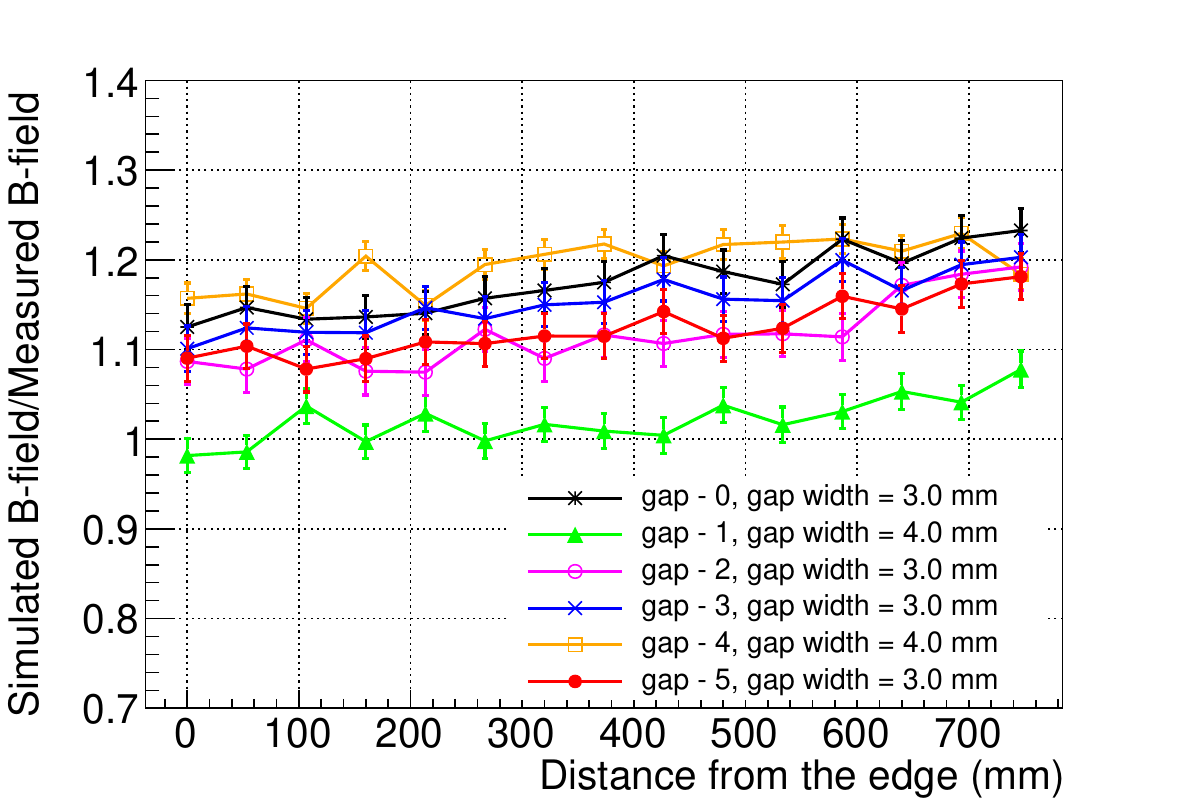}
    \includegraphics[width=0.49\textwidth]{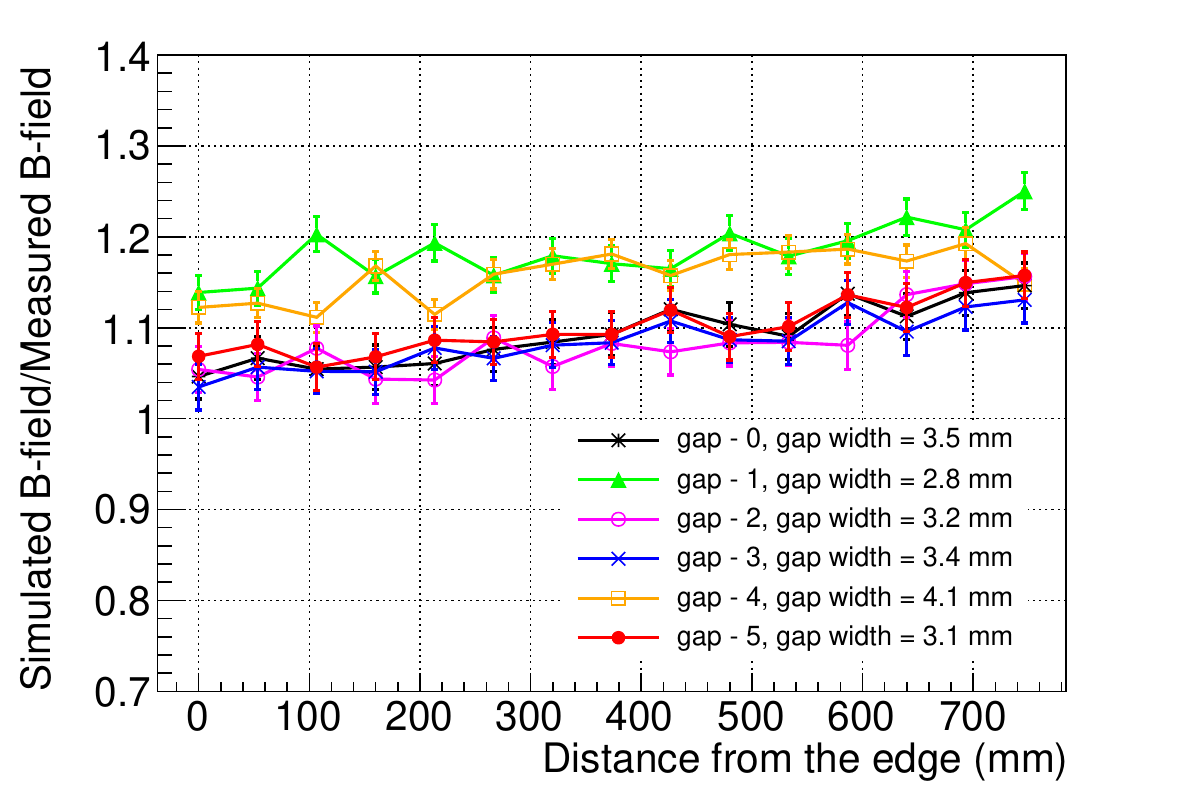}
\caption{As in Fig.~\ref{fig:sim_vs_meas_500_amps} for the coil current
of 700 A; The ratio for both sets of gaps (0, 2, 3 and 5) and (1,4)
now exceeds unity for all gaps.}
    \label{fig:sim_vs_meas_700_amps}
\end{figure}

\begin{figure}[htp]
  \centering
    \includegraphics[width=0.49\textwidth]{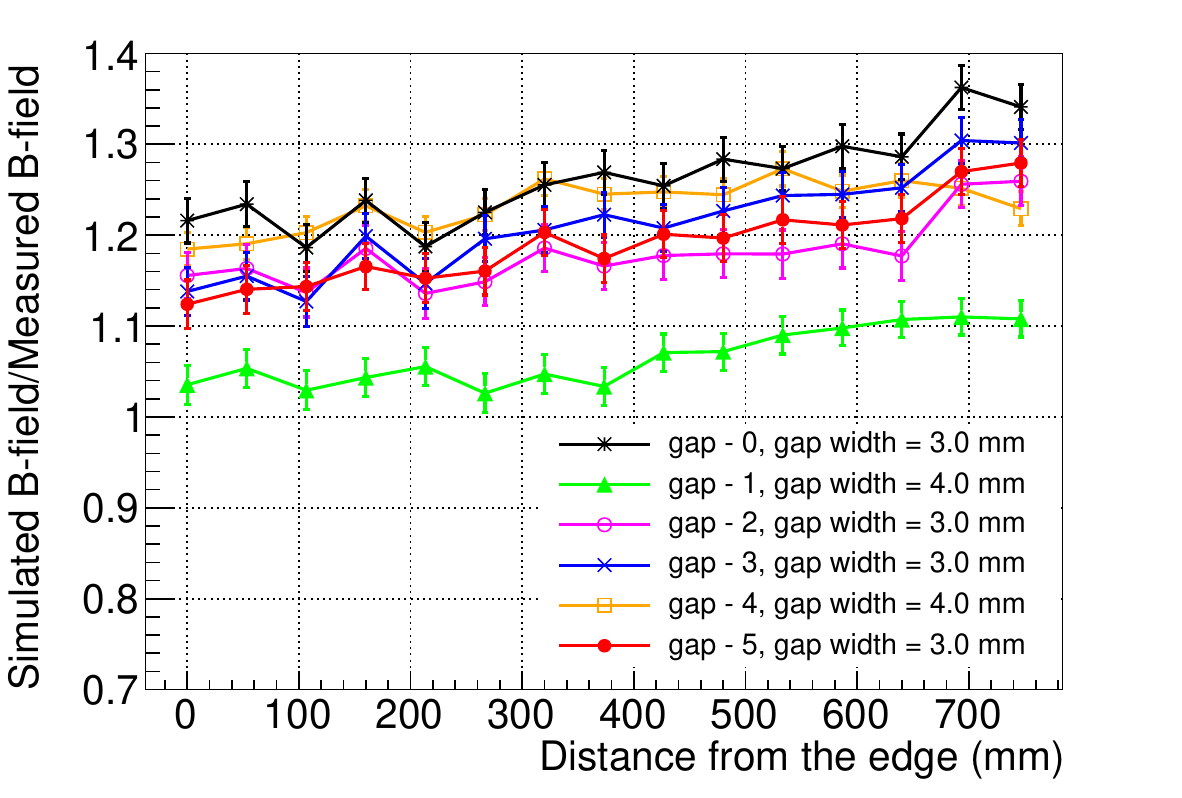}
    \includegraphics[width=0.49\textwidth]{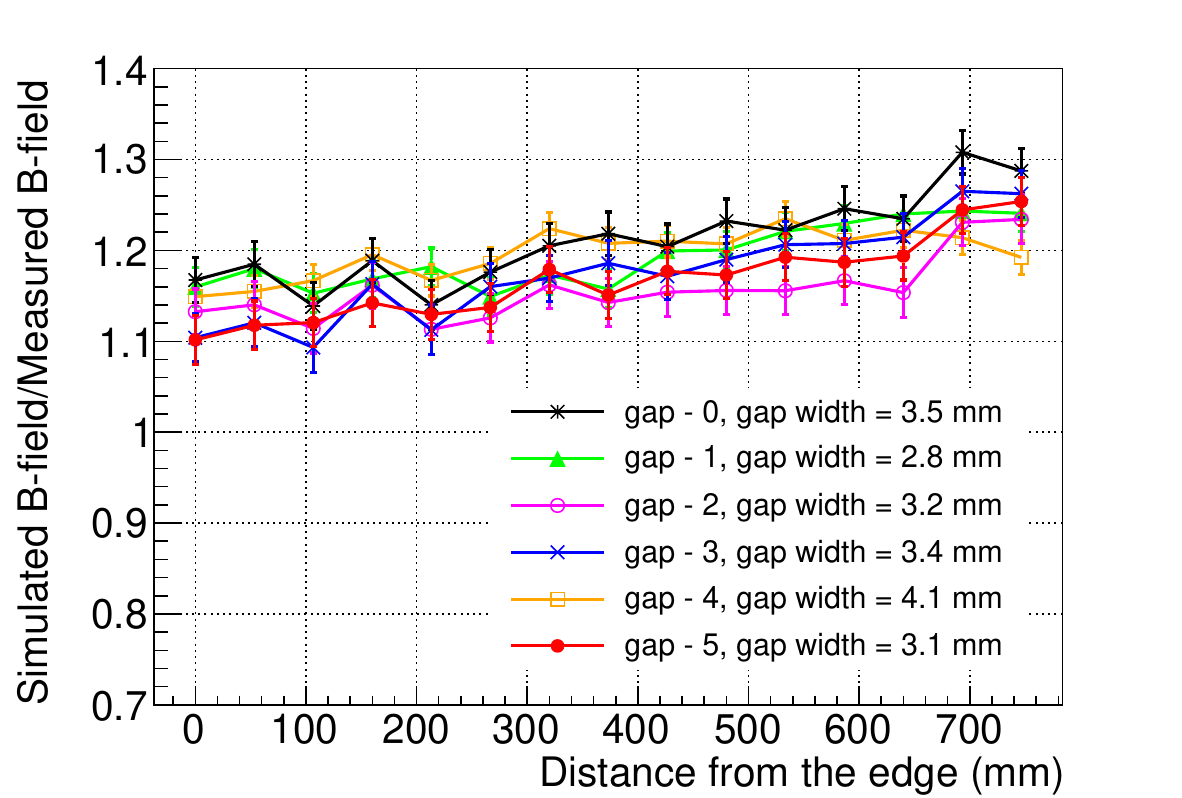}
\caption{As in Fig.~\ref{fig:sim_vs_meas_500_amps} for the coil current
of 900 A; the ratio for both sets of gaps (0, 2, 3 and 5) and (1,4)
exceeds unity for all gaps, but they are also converging to the same
value for all six gaps.}
    \label{fig:sim_vs_meas_900_amps}
\end{figure}

\paragraph{Coil current 500A}: If the measured magnetic field is
compared with the simulations using an ideal gap value (i.e., 3 mm
for gaps 0, 2, 3, 5 and 4 mm for gaps 1 and 4) and ratio of simulated
magnetic field vs measured magnetic field is plotted, it can be seen
from left of Figs.~\ref{fig:sim_vs_meas_500_amps} (for coil current
of 500 A), ~\ref{fig:sim_vs_meas_700_amps} (for coil current of 700
A) and ~\ref{fig:sim_vs_meas_900_amps} (for coil current of 500 A)
that the results for gaps 2 and 5 are similar, indicating that their
gap widths are similar, which is indeed the case as can be seen from
Table~\ref{tab:gap_width_new}. The measured magnetic field in the  gap-3
is more than gap-0 as can be seen that gap width is more for gap-0
among gaps - 0, 2, 3 and 5.  Also, since the ratios for gaps 0 and 3
are larger than for gaps 2, 5 we expect that the former have larger gaps
than the latter, which is also true. Finally, we expect the ratios for
gaps 1 and 4 to be similar but they are very different, again due to
the considerable difference in their gap widths. Hence we see that the
differences in measured values are significantly driven by the gap widths.

If the measured gap width (with gap width correction) is taken into
account, the measured values of the magnetic field (and hence the
simulated to measured ratios) are in agreement to within $\pm$ 5\% of each
other (see right side of Fig.~\ref{fig:sim_vs_meas_500_amps}) for the gaps
0, 2, 3 and 5 after gap correction and gaps 1 and 4 also. At 500 A coil
current, the group of gaps-3 mm (0, 2, 3 and 5) are bunched together and 4
mm (1 and 4) are bunched together separately with a difference of $\sim$
10\%, indicating a mis-match between the simulations in the peripheral
and side regions. This may happen because at coil current of 500 A,
we are still in the linear region in the $B$--$H$ curve and any deviation
in the B-H curve or an inhomogeneity in the bulk of the iron plate
could be responsible for the deviation. This is particularly so since we
have already seen, both in the measurement and the simulations, that the
field around the horizontal gaps saturates at much higher currents that
the vertical ones.

\paragraph{Coil current 700 A}: As the coil current is raised further
towards saturation from 500 A to 700 A, the ratios of simulated to
measured field for both the groups of gaps (0, 2, 3 and 5) and (1 and 4)
begin to come closer together (see right side of
Fig.~\ref{fig:sim_vs_meas_700_amps}). However, the ratio for both sets
now exceeds unity for all gaps.

\paragraph{Coil current 900 A}: This trend continues for coil
current of 900 A: the ratios of simulated to measured field after
gap-width correction begin to merge with each other (see right
Fig.~\ref{fig:sim_vs_meas_900_amps}) for all six gaps, both the horizontal
and vertical ones. Two points may be observed here: the simulated
magnetic field is larger than the measured one for all gaps over their
entire length. In addition, the ratio has an increasing slope,
indicating that the field simulated inside the iron, near the coils is
much larger than the measured one. This slope is especially perceptible
for gaps 0,2,3,5, so that the ratio increases by 15\% from the outer
edge to the centre. The slope is much smaller for gaps 1,4 where the
corresponding ratio increases by only 7\%.

\section{Results and Discussion}

A few remarks are in order. It is important to note that the field is
fairly uniform over the entire gap width. Hence the thickness of the
Hall probe, and whether the sensor was mounted on one side or other
of the probe, does not change the measurement; otherwise it would be
difficult to make a meaningful statement about the $B$-field in the
gaps. In addition, while the field may differ substantially in the gaps
depending on the width, the field in the iron itself remains practically
the same throughout, for a given current, independent of the gap width
(except at locations just adjacent to the gaps). Hence, once the field
is measured in the gaps and the simulations result matched with these,
it is not necessary to measure the fields at {\em all} gaps; the magnetic
field map can be validated by just a few sample measurements. This is
important, since it is not possible to measure the gap widths in all
layers. Measuring the gap widths at the time of assembly is also not
sufficient since they are different with and without the field due to
torsional forces that bring the plates together when the current is
switched on.

It is clear that the gap thickness measurement plays a crucial role while
comparing the simulated and measured magnetic field. It was shown in
Ref.~\cite{Khindri:2023anm} while studying the physics reach of ICAL,
that while local random fluctuations of the magnetic field could be
well-tolerated, systematic deviations from the true value (such as have
been observed in this study) will significantly impact the results. For
example, this may cause the fits to converge to a best-fit value different
from the true/input value of the neutrino oscillation parameters such
as $\sin^2\theta_{23}$ and $\Delta m^2_{32}$.

The present study shows that as the coil current approaches 900 A starting from 500 A,
the simulated magnetic field is more than the measured
magnetic field as can be seen from the right hand figures of
Figs.~\ref{fig:sim_vs_meas_500_amps}, \ref{fig:sim_vs_meas_700_amps}
and \ref{fig:sim_vs_meas_900_amps}. In addition, while the magnetic field
in the gaps 0, 2, 3, 5 increases from the edges towards the center of the
detector, this increase is more pronounced in the simulations
(especially for the vertical gaps) as compared
to measurement, as discussed above. The origin of this discrepancy could
be due to differences in the $B$-$H$ curve used in the simulations and
the actual performance. Notice from Fig.~\ref{fig:B_H_curve} that
different samples gave very similar $B$-$H$ curves. Hence, this is
unlikely to be the source of this discrepancy. However, the plate
manufacturing procedure that was used to cut the plates can generate
heat-affected zones (HAZ) where the magnetic field properties can be
significantly altered. Inhomogeneities in the bulk low-carbon steel
plates and other material unaccounted for in the simulations (the
simulations includes only the iron and copper coils and not the RPCs, 
electronics, etc) can also substantially degrade the magnetic field from
the expected value. Minimizing the heat affected zone (HAZ) may help to
improve the agreement with the simulations. However, the measurement can
be used to normalize and calibrate the simulations in the short term,
and guide other measurements that might help pinpoint the reasons for
the discrepancy and help in a better understanding of the magnetic field in
mini-ICAL. Although outside the scope of the present work, measurements
on an appropriate prototype magnet using secondary beams of muons of 
known momentum would help validate the entire procedure.

\appendix
\section{Simulations with different number of layers}
\label{app:a}

As described in Section \ref{sec:no_of_layers}, the number of iron layers
used in the model affects the simulated magnetic field. To confirm this,
a study is done using models with different number of layers. Here, the
dimensions of C and D plates is changed from 1200 $\times$ 2000 mm and
1962 $\times$ 1600 mm to 1200 $\times$ 3000 mm and 2962 $\times$ 1600
mm to check if the change in geometry makes any difference. This study
is done for 900 A coil current with default mesh since it was done just
to check the trend.

It is observed that the number of layers present in
the model as well as their position with respect to each other makes a
difference in the output magnetic field. We explain this as follows. We
have considered 5 different models in addition to the standard 11-layer
model. Two of them have 3 layers, placed at positions (1,6,11) or
(5,6,7) of the 11-layer model. Two of them have 5 layers, placed at
positions (1,3,6,9,11) or (2,4,6,8,10) of the 11-layer model. Finally,
one model has just a single layer, placed in position 6 (centre) of the
11-layer model.

\begin{figure}[htp]
 \centering
\includegraphics[width=0.85\textwidth]{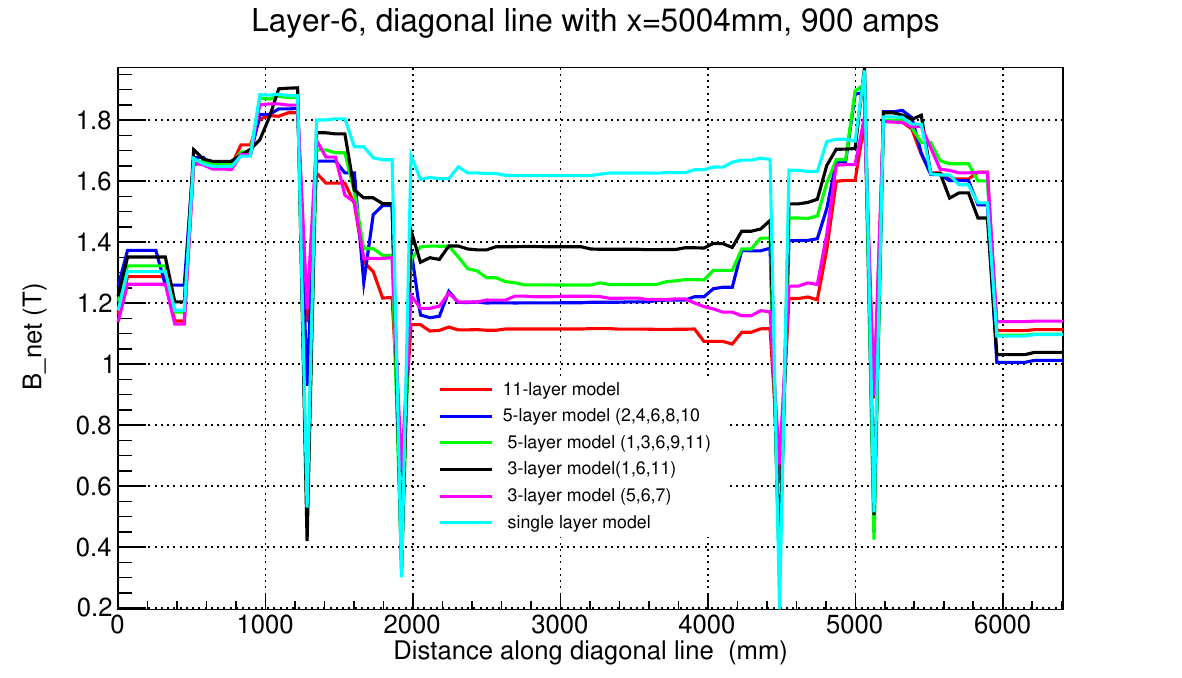}
\caption{Magnetic field along the diagonal line $x=y$ in the central layer for
models with different number of layers and different relative positions
of the layers, as shown.}
\label{B_diag_layers}
\end{figure}

In Fig.~\ref{B_diag_layers} the magnetic field value along the
diagonal line is shown for the middle layer (layer-6) for all 6 models
considered. It can be seen that the magnetic field in this central
layer is maximum for the single layer model and minimum for the
11-layer model. As 3 or 5 layers are added, the field decreases from
the 1-layer model and is intermediate  between the 1-layer and 11-layer
models. However, the {\em relative placement} of the layers also matters:
the field in the central module is always smaller either if there are
more layers (2,4,6,8,10) or if the adjacent layers are closer to the
central layer (5,6,7). Hence the 5-layer model (1,3,6,9,11) shows a larger
magnetic field in layer-6 than the 5 layer model (2,4,6,8,10) where the
layers adjoining the central one are closer.  Similarly, the 3-layer
model (1,6,11) with very distant neighbours has a larger magnetic field
in layer 6 than the 3-layer model (5,6,7) with closely adjacent layers.
Clearly the magnetic field in a layer depends on the number of adjoining
layers as well as their relative distances from that layer.
Finally, it may be noted that the behaviour is roughly inverted in the region
outside the coil slots.

\acknowledgments
We sincerely thank the entire INO collaboration for their support. In particular, 
we would like to appreciate Mr. K.C. Ravindran and Dr. Umesh L. as well as Mr. 
Hritik Gogoi and Mr. Debasish Gayan  (Cotton University, Guwahati) for their 
help during the measurements on the mini-ICAL detector at IICHEP, Madurai. 
Initial work by Annabelle Dani, Monika Budania, Sanika Itagi and Lawansh 
Singh -- all then the students of Fr. C.R.I.T, Mumbai in developing the Hall 
sensor probes is gratefully acknowledged. Thanks are also due to Mr. K.V. 
Thulasi Ram (TIFR, Mumbai) for his help during the calibration of the probes.


\end{document}